\documentclass[superscriptaddress,11pt]{article}
\usepackage{authblk}
\usepackage{array}
\usepackage{type1cm}
\usepackage{lettrine}
\usepackage{graphicx}
\usepackage{geometry}
\usepackage{psfrag}
\usepackage{amsmath}
\usepackage{amsfonts}
\usepackage{appendix}
\usepackage[margin=10pt,font=footnotesize,labelfont=sc,format=hang,labelformat=parens]%
{caption}
\usepackage{subcaption}
\usepackage{float}
\usepackage{subcaption}
\usepackage{accents}
\usepackage{amssymb}%
\providecommand{\U}[1]{\protect\rule{.1in}{.1in}}

\numberwithin{equation}{section}
\def\be{\begin{equation}}
\def\ee{\end{equation}}
\def\ba{\begin{eqnarray}}
\def\ea{\end{eqnarray}}
\def\bi{\begin{itemize}}
\def\ei{\end{itemize}}

\def\brho{\bar{\rho}}
\def\u{\bar u}
\def\I{\cal I}
\def\g2{{}^{(2)}g}
\def\fg{{}^{(4)}g}
\def\2R{{}^{(2)}  {\cal R}  }
\def\4R{{}^{(4)}   {\cal R}   }

\def\bra{\langle}
\def\ket{\rangle}

\def\e{\epsilon}

\def\fp{f_{(+)}}
\def\fm{f_{(-)}}

\begin{document}

\title{Spherical collapse and black hole evaporation}

\author{Madhavan Varadarajan}

\affil{Raman Research Institute\\Bangalore-560 080, India}
\maketitle

\begin{abstract}

We consider spherically symmetric gravity coupled to a spherically symmetric scalar field with a specific coupling which 
depends on the Areal Radius. Appropriate to spherical collapse, we require the existence of an
axis of symmetry
and consequently a single asymptotic past and future (rather than a pair of `left' and `right' ones).
The scalar field stress energy takes the form of null dust. 
Its classical collapse is described by the  Vaidya solution. From a two dimensional `$(r,t)$' perspective, the scalar field is conformally coupled so that its quantum
stress energy expectation value is well defined. Quantum back reaction is then incorporated through an explicit formulation of the 4d semiclassical  Einstein equations.
The semiclassical 
solution describes black hole formation together with its subsequent evaporation along a {\em timelike} dynamical horizon 
(i.e. a timelike outer marginally trapped `tube').
A balance law at future null infinity relates the rate of change of  a back reaction-corrected Bondi mass to a  manifestly {\em positive} flux.
The detailed form of this balance law together with a proposal for the dynamics of the true degrees of freedom underlying the
putative non-perturbative quantum gravity theory is supportive of the paradigm of singularity
resolution and information recovery proposed by Ashtekar and Bojowald.  In particular all the information
including that in the collapsing matter is expected, in our proposed scenario, to emerge along a
single ‘quantum extended’ future null infinity. 
Our  
analysis is  on the one hand  supported and informed by earlier numerical work of Lowe \cite{lowe} 
and Parentani and Piran \cite{pp}
and on the other, serves to clarify certain aspects of their work through our explicit requirement of the existence of an axis of symmetry.
\end{abstract}

\section{Introduction}
The aim of this work is to study black hole evaporation and the Hawking Information Loss Problem \cite{hawking}
in the simplified context of spherical symmetry. We are interested in the general relativistic spherical
collapse of a matter field in a context which allows analytical understanding of its classical collapse to a black hole as well the computation of its quantum back reaction on the collapsing spacetime geometry.
These aims are achieved by choosing the matter field to be a spherically symmetric massless scalar field and by defining  its coupling to gravity to be a modification of standard minimal coupling, this modification being  dependent on the areal radius of the spheres which comprise the orbits of the angular Killing fields of the spacetime.

We require the spacetime to be asymptotically flat  in the distant past. 
Initial data for the matter field is specified at past null infinity. This matter then collapses to form a black hole.
Since the 4d spacetime is spherically symmetric and in the distant past looks like standard 4d Minkowski
spacetime,  the axis of symmetry is located {\em within} the spacetime 
as a 1d line  which is timelike in the 
distant past. We restrict attention to the case wherein the 
axis of symmetry is timelike everywhere. By definition, the Areal Radius $R$ vanishes along the axis.
We note here that the axis is distinguished from the $R=0$ classical black hole black hole singularity in that the 
spacetime geometry at the axis is {\em non-singular}.

The existence of the axis of symmetry in  collapsing spherically symmetric spacetimes is a key feature which differentiates  such spacetimes from those of   eternal black holes. In the case of spherical symmetry, the eternal black hole
geometry is that of the  Kruskal extension of Schwarzschild spacetime.  Such an eternal black (and white) hole spacetime does not have an axis of symmetry;
instead, and in contrast to a collapse situation, this spacetime has not 
one, but {\em two} sets of infinities, a left set and a right set.

The analysis of information loss  in the context of spacetimes with a pair of infinities becomes problematic
as can be seen in the example of the 2d CGHS model.  In this model \cite{atv}, the vacuum
state at {\em left} past null infinity is viewed as Hawking radiation by observers at
at {\em right} future null infinity,  whereas the collapsing matter (and, hence,  the information
regarding its nature) propagates from {\em right} past null infinity towards {\em left} future null infinity.
While much has been learnt about black hole evaporation and the purity of the state along an anticipated quantum
extension of right future null infinity 
in this model \cite{atv}, one of its unsatisfactory features is the existence of the pair of left and right infinities.
\footnote{\label{fn1/m}Another is the mass independence of the Hawking temperature which is a consequence of 
the detailed horizon red shift being different from the 4d gravitational one. Since the spacetime geometry
studied in this work is general relativistic, the horizon red shift is the standard one as is  the inverse mass dependence of the Hawking temperature.}
From this point of view,  the importance of  the incorporation of the axis of symmetry in the spacetimes considered in this work is that, as a consequence, such spacetimes have a {\em single} set of infinities.

A description of the results obtained in this paper together with its layout  is as follows. In section \ref{sec2}, we discuss the kinematics of spherical symmetry. We describe
the coordinates used, the location of the axis in these coordinates and discuss the behavior of fields at the axis. We prescribe
`initial'  conditions  for the geometry which ensure asymptotic flatness in the distant past and as well as for the nature of matter data in the distant past. 
In section \ref{sec3} we describe the classical dynamics of the
system. We exhibit the action and show that the matter stress energy takes the form of a pair of (infalling and outgoing) streams of null dust.  We show that the 
dynamical equations together with  the initial conditions and the requirement of axis existence are solved by the Vaidya spacetime 
(in which  collapse of the {\em infalling} null dust stream forms a black hole).
In section \ref{sec4} we derive the semiclassical equations which incorporate back reaction and then  
combine analytical results and  physical arguments with prior numerical work to describe the geometry of the semiclassical solution.
This geometry corresponds to the formation of a black hole through spherical collapse of the scalar field and its subsequent evaporation through quantum radiation
of the scalar field. 
In section \ref{sec5} we analyse the semiclassical equations in the distant future and show that they imply
a balance law relating the {\em decrease} of a quantum backreaction corrected Bondi mass to a {\em positive}, back reaction
corrected Bondi flux. We argue that the detailed nature of this balance law suggests, in a well defined manner,  that the classical future null infinity 
admits a quantum extension wherein correlations with the Hawking radiation
manifest so that the state on this extended future null infinity is pure.

In section \ref{sec6} we combine the results of sections \ref{sec4} and \ref{sec5} together with informed speculation
on the nature of the true degrees of freedom of the system at the deep quantum gravitational level and thereby propose
a spacetime picture which encapsulates  a possible solution of the Information Loss Problem. The solution
is along the lines of the Ashtekar-Bojowald paradigm \cite{ab} wherein quantum gravitational effects resolve the classical black hole singularity opening up a vast quantum extension of classical spacetime beyond the hitherto  classically singular region wherein correlations with the Hawking radiation and information about the collapsing matter emerge. Section \ref{sec7} is devoted to a discussion of our results and further work. Some technical details and proofs are collected in Appendices.

In what follows we choose units in which $c=1$. We shall further tailor our choice of units in section \ref{sec3} so as to set certain coupling constants
to unity.

\section{\label{sec2} Kinematics in spherical symmetry}
\subsection{\label{sec2a}Spacetime geometry}
Choosing angular variables along the rotational killing fields, the spherically symmetric line element takes the form:
\be
ds^2= {\g2}_{\mu \nu}dx^{\mu} dx^{\nu} +  R^2 (d\Omega)^2,\;\; \mu , \nu =1,2.
\label{4gsph}
\ee
Here $R$ is the areal radius and $(d\Omega)^2$ is the line element on the unit round 2 sphere which in
polar coordinates $(\theta, \phi)$ is $(d\theta)^2 + \sin^2\theta (d\phi)^2$.  The space of orbits of the rotational killing fields is 2 dimensional.
The pull back of the 4- metric to this 2 dimensional `radial-time' space is the Lorentzian 2-metric $\g2$.
The areal radius $R$ depends only on coordinates on this 2d spacetime and not on the angular variables.
Choosing these coordinates  $\{x^{\mu}\}$ to be along the radial outgoing and 
ingoing light rays and denoting these coordinates by $(x^+,x^-)$ puts the 2-metric in conformally flat form:
\be
\g2_{\mu \nu}dx^{\mu}dx^{\nu} = -e^{2\rho}dx^+ dx^- = e^{2\rho}(-(dt)^2 + (dx)^2)
\label{g2pm}
\ee
where we have set 
\be
x^{\pm}= t\pm x
\label{pmtx}
\ee
The areal radius $R$ is a function only of $(x^+, x^-)$.  The area of a spherical light front at
fixed $x^+,x^-$ is  {$4\pi R^2$}. Hence, outgoing/ingoing expansions of spherical light fronts 
are proportional to {$\partial_+R, \partial_-R$}. In particular, a spherical  outer marginally trapped surface 
located at fixed $x^+,x^-$
is defined by the conditions
\be
\partial_+R= 0, \;\;\;\;\;   \partial_-R <0
\label{marginal}
\ee
For future reference we note here that an outer marginally trapped tube formed by a 1 parameter family of outer marginally trapped surfaces 
is referred to as a {\em dynamical horizon}. The notion of a dynamical horizon was originally introduced in \cite{aadh} and refined in \cite{aapersp} 
(we use the latter definition here).

As indicated in the Introduction we restrict attention to the case in which the axis of symmetry is a timelike curve located within the  4d spacetime. Hence the axis is located at $x^+=F(x^-)$, with 
$\frac{dF}{dx^-} >0$. By using the conformal freedom available in the choice of our conformal coordinates, we can choose $F(x^-)$ to be our new $x^-$ coordinate. With this  choice, the axis
is located along the straight line
\be
x^+=x^- \equiv x=0.
\label{axislocation}
\ee
Next, we require that the 4- metric is asymptotically flat as $x^- \rightarrow {-\infty}$ so that past null infinity, $\I^-$,  is located  at $x^- = -\infty$. In conformal coordinates the detailed fall off conditions
at $\I^-$
turn out to be: 
\be
R = \frac{x^+- x^-}{2} + O(1/x^-)\;\; \;\;\;\;   e^{2\rho}= 1 + O(1/(x^{-})^2)
\label{aflat}
\ee
As we shall see in section \ref{sec3}, the Vaidya solution satisfies these conditions and in this solution the mass information is 
contained in the $O(1/x^-)$ part of $R$ and the 
$O(1/(x^{-})^2)$ part of $e^{2\rho}$.   

It is straightforward to see that (\ref{axislocation}) and (\ref{aflat})  fix the conformal freedom in the choice of the $x^{\pm}$ coordinates upto (the same) 
constant translation $c$ i.e. $x^{\pm}\rightarrow x^{\pm} + c$.
\footnote{
Let $X^{\pm}(x^{\pm})$ be new conformal coordinates. Equation (\ref{axislocation}) and invertibility of $X^{\pm}(x^{\pm})$
imply $X^{+}$ and $X^{-}$ are identical functions of their arguments. The coordinate location of $\I^-$ implies that
$X^{-}(x^-\rightarrow -\infty)\rightarrow -\infty$.
The first equation of (\ref{aflat}) implies 
$\frac{x^+- x^-}{2} + O(1/x^-)= \frac{X^+(x^+)- X^-(x^-)}{2} + O(1/X^-(x^-))$; the difference of evaluations of this equality at $\I^-$,  for  arbitrary $x^+$ 
and for some fixed $x^+_1$ yields $X^+(x^+)= x^+ + (X^+(x^+_1)-x^+_1)$ which proves the desired result.
}
As we shall see in section \ref{sec7}, our results 
are independent of this remaining choice of coordinates and we fix them once and for all hereon.


To summarise: The region of interest for us in this paper is the $x\geq 0$ part of the Minkowskian
plane with $\I^-$ located at $x^-=- \infty$ and the axis at $x=0$.
 Each point  $(t,x)$ on this half plane represents a 2 sphere of area $4\pi R^2(t,x)$ with 
$R$ vanishing along the axis of symmetry at  $x=0$, this axis serving as a boundary of the region of 
interest. 

Finally, recall from the Introduction that the axis with $R=0$ is distinguished from the expected classical
singularity at $R=0$ by virtue of the geometry at the axis being non-singular. In the specific classical and semiclassical spacetime solutions
which we study in this work, the geometry in a neighborhood of the axis will turn out to be flat (and hence non-singular).
The physical spacetime geometry in these solutions will occupy a subset of the half $(t,x)$ plane due to the occurrence of singularities (in the classical
and semiclassical solutions) or Cauchy horizons (in the semiclassical solutions). For details see sections \ref{sec3},\ref{sec4}.

\subsection{\label{secmkin}Matter}
The matter is a spherical symmetric scalar field $f (t,x)$.
Note that at the axis, $R=0$ so that $\partial_t R=0$ there. The requirement that the geometry near the axis is non-singular together with the assumption that 
$x^{\pm}$ are good coordinates for the 2d geometry implies that 
$\partial_xR=e^{\rho}$ at the axis (see Appendix \ref{seca1} for details).
This ensures that at the axis $(t,R)$ is a good chart. 

Recall that the axis is a line in the 4d spacetime. We require that $f$ be differentiable at the axis
from this 4d perspective.  In particular consider differentiability along a $t$= constant  radial line 
which starts out at, say, $R>0$ and moves towards the axis along a trajectory
which decreases $R$. Once it moves through the axis, $R$ starts increasing again.
Differentiability of $f$ at the axis then demands that 
$-\frac{\partial f}{\partial R}|_{R=0} = +\frac{\partial f}{\partial R}|_{R=0}$
which in turn implies that $\frac{\partial f}{\partial R}|_{R=0}=0$.
Reverting to the $(t,x)$ coordinates this implies that 
\be
\frac{\partial f}{\partial x}|_{t,x=0}=0
\ee
which in $(x^+, x^-)$ coordinates takes the `reflecting boundary condition' form at the axis:
\be
\frac{\partial f}{\partial x^+}|_{t,x=0}= 
\frac{\partial f}{\partial x^-}|_{t,x=0}
\label{reflect}
\ee
In addition to these boundary conditions we demand that $f$ be of compact support
on $\I^-$. Finally, we require that $f$   satisfies  the following condition at $\I^-$.
Define 
\be
\frac{1}{2}\int_{x^+_i}^{x^+}d{\bar x}^+(\partial_+ f({\bar x}^+, x^-\rightarrow -\infty))^2 = m(x^+),
\label{defm}
\ee
where $f$ is supported between $x^+_i$ and $x^+_f >x^+_i$ on $\I^-$.
We require that $f$ be such that: 
\be
\lim_{x^+\rightarrow (x^+_i)^+}\frac{m(x^+)}{x^+- x^+_i} >\frac{1}{16}
\label{pc}
\ee
where the limit is to be taken as $x^+$ approaches $x^+_i$ from the right (i.e. $x^+>x^+_i$).
As we shall see, condition (\ref{pc}) ensures that the prompt collapse Vaidya spacetime is a classical solution (by the  prompt collapse Vaidya spacetime we mean
one in which the singularity is neither locally nor globally naked \cite{bengtsson}).

\section{\label{sec3} Classical dynamics}
\subsection{\label{action}Action}
The action for the spherically symmetric 4-metric $\fg$ is the Einstein Hilbert action:
\be
S_{geometry}= \frac{1}{8\pi G}\int d^4x\sqrt{-\fg} {\4R}
\label{sgeo}
\ee
Assuming spherical symmetry, integrating over angles and dropping total derivative terms (in our analysis we have ignore the issue of the addition of suitable boundary terms
to (\ref{sgeo}) so as to render the action differentiable), we obtain:
\be
S_{geometry}= 
\frac{1}{2G}\int d^2x\sqrt{-\g2}R^2[\2R+ 2(\frac{\nabla R}{R})^2 +2R^{-2}]
\label{sgeosph}
\ee
The matter coupling is chosen to depend on the Areal Radius $R$ so that the matter action is:\\
\be
S_{matter}= -\frac{1}{8\pi}\int d^4x\sqrt{-\fg} {\fg}^{\hspace{-0.25mm}ab}{ \frac{1}{R^2}  } ({\nabla_a f\nabla_b f})
\label{smatter}
\ee
where $f$ is spherically symmetric and hence angle independent.
Integrating over angles, we obtain:
\be
S_{matter}= -\frac{1}{2}\int d^2x \sqrt{-\g2} (\nabla f)^2
\label{sfsph}
\ee
so that the Areal Radius dependent  4d coupling of $f$ to the metric $\fg$  in (\ref{smatter})  reduces to 2d conformal coupling to the metric $\g2$
in (\ref{sfsph}).
The total action is then:
\ba
S&=&S_{geometry}+ S_{matter}\nonumber \\
&=&\frac{1}{2G}\int d^2x\sqrt{-\g2}R^2[\2R+ 2(\frac{\nabla R}{R})^2 +2R^{-2}]
 -\frac{1}{2}\int d^2x \sqrt{-\g2} (\nabla f)^2 
\label{ssph}
\ea
To summarise: The action for the geometry is exactly that of General Relativity reduced to the spherical symmetric sector. This ensures that in classical
black hole solutions the infinite redshifting of light which propagates along the event horizon to $\I^+$ is exactly as in general relativity. This in turn
ensures that the Hawking temperature in a QFT on CS calculation is the standard one with inverse mass dependence (see Footnote \ref{fn1/m} and remarks in section 
\ref{secfhat}). In contrast the matter action does not arise through a spherically symmetric reduction of a 4d covariant action. In particular it differs from
such a reduction of minimal coupling due to the Areal Radius dependence of the coupling; since the Areal Radius is defined only in the spherically 
symmetric setting, the matter action is defined only in the  spherically symmetric context. However this is a small price to pay for the classical solvability
of the resulting equations (see section \ref{secvaidya}) and the explicit computation of back reaction (see section \ref{sec4}).
As we shall see in sections \ref{seceom}, \ref{secvaidya}  the 2d conformal coupling in (\ref{sfsph}) results in a matter dynamics where there is no backscattering of the classical scalar field
off the curvature; this is in contrast to the minimally coupled spherically symmetric case where such backscattering renders the dynamics non-amenable to
analytic solution. As we shall see explicitly in section \ref{secfhat} the 2d conformal coupling allows the application of the classic results of Davies
and Fulling \cite{fd} towards the well-definedness of the matter stress energy expectation value and the explicit formulation of the semiclassical Einstein equations.
\\


\noindent{\bf Note:} Since the dynamics in spherical symmetry is effectively 2 dimensional, it may also be viewed as a 2 dimensional dilatonic gravity system.
In order to make contact with this view, we define  a `dimensionless areal radius' ${\bar R}$ as 
${\bar R}:= \kappa^2 R^2$ with $\kappa$ being an arbitrarily chosen (but fixed) constant with dimensions of inverse length.
In the 2d gravity literature $\kappa$ is often referred to as a `cosmological constant'. 
It is then straightforward to check that the action (\ref{ssph}) takes the form:
\ba
S&=&S_{geometry}+ S_{matter}\nonumber \\
&=&\frac{1}{2G\kappa^2}\int d^2x\sqrt{-\g2}{\bar R}^2[\2R+ 2(\frac{\nabla {\bar R}}{{\bar R}})^2 +2 {\bar R}^{-2}\kappa^2]
 -\frac{1}{2}\int d^2x \sqrt{-\g2} (\nabla f)^2 . 
\label{ssphk}
\ea
The interested reader may  further substitute ${{\bar R}^2}=: e^{-2\phi}$ in the above action to obtain one in terms of the 2d gravitional
metric $\g2$, the matter field $f$ and the dilaton $\phi$. 
Finally, note that in addition to setting $c=1$ we may choose units such that $\kappa =1$.
With this choice $R={\bar R}$ and there is no difference between the form of the actions (\ref{ssphk}) and (\ref{ssph}). 
Hence, in what follows, we shall work with (\ref{ssph}) and the reader may simply ignore the note above or take the 2d dilatonic gravity view in units 
in which $\kappa =1$. As we shall see below, in addition to $c=1$ we shall also set $G=1$ by choice of units but {\em shall not} set $\hbar=1$. From the point of view of
dilatonic gravity this implies fixing units such that $\kappa=G=c=1$, where we recall that $\kappa$ was chosen arbitrarily. On the other hand if 
we do not take the dilatonic gravity view,  we do not  introduce an arbitrarily chosen $\kappa$ and  we only set $G=c=1$.

\subsection{\label{seceom}Dynamical equations}
In what follows we shall often employ the obvious notation
$\frac{\partial A}{\partial x^{\pm}}\equiv \partial_{\pm}A \equiv A,_{\pm}$ for partial derivatives of a function $A$.
Also, by  $G_{{\hat \Omega} {\hat \Omega}}$ below we mean the component $G_{ab}{\hat \Omega}^a {\hat \Omega}^b$ of the tensor $G_{ab}$ 
where  ${\hat \Omega}^a$ is a unit vector tangent in the direction
of a rotational Killing vector field (for e.g. in polar coordinates we could choose ${\hat \Omega}^a = R^{-1} (\frac{\partial}{\partial \theta})^a$).

Finally, we shall use units in which , in addition to our choice of $c=1$ made at the end of the Introduction, we also set $G=1$. We shall however
explicitly retain factors of $\hbar$ so we {\em do not} set $\hbar =1$.

Since the action for the geometry is the Einstein-Hilbert action, the equations of motion which follow from (\ref{ssph}) are just the Einstein
equations for a spherically symmetric metric coupled to the matter field $f$ i.e. the equations take the form
$G_{ab}= 8 \pi T_{ab}$ where $G_{ab}$ is the Einstein tensor for the spherical symmetric 4 metric $\fg$ (\ref{4gsph}), (\ref{g2pm}):
\ba
-\frac{e^{2\rho}}{4}
G_{{\hat \Omega} {\hat \Omega}}&=& \partial_+\partial_- \rho + \frac{1}{R}\partial_+\partial_-R=0 \label{gangle} \\
R^2\;G_{+-} &=  &  2R\partial_+\partial_-R + 2 \partial_+R\partial_-R + \frac{1}{2} e^{2\rho}=0  \label{gpm} \\
R^2\;G_{\pm \pm} &= & R^2[-\frac{2}{R}( \partial_{\pm}^2R- 2\partial_{\pm}\rho \partial_{\pm}R)]= (\partial_{\pm}f)^2
\label{gppmm}
\ea
The remaining components of the Einstein tensor vanish as a result of spherical symmetry. 
From (\ref{gangle})- (\ref{gppmm}),
the only non-vanishing components of $T_{ab}$ are 
\be
T_{\pm\pm} = \frac{1}{4\pi R^2}\frac{(\partial_{\pm}f)^2}{2}.
\label{tvaidya}
\ee
Since the matter is conformally coupled, it satisfies the free wave equation on the fiducial flat $x^+,x^-$ spacetime. Explicitly, varying $f$ in the action
(\ref{ssph}) yields:
\be
\partial_{+}\partial_{-}f =0
\label{feq}
\ee

\subsection{\label{secvaidya}Classical Solution: Vaidya spactime}

Since $f$ satisfies the free 1+1 wave equation on the fiducial flat spacetime, solutions take the form of the sum of left and right movers:
\be
f(x^+,x^-) = \fp (x^+) + \fm(x^-). 
\label{fpm}
\ee
Since the solution (\ref{fpm}) has to satisfy reflecting boundary conditions (\ref{reflect}) at the axis, it follows that:
\be
\partial_+{\fp}(x^+)|_{x^+=t} = \partial_-(\fm(x^-)|_{x^-=t}    \;\;\forall t
\label{fpmbc}
\ee
We shall restrict attention to $f_{\pm}$ of compact support in their arguments. Equation (\ref{fpmbc}) then implies that:
\be
\fp (y)= \fm (y)  \;\;\; \forall y
\label{fp=fm}
\ee

\begin{figure}
\begin{center}
\includegraphics[width= 0.5\textwidth]{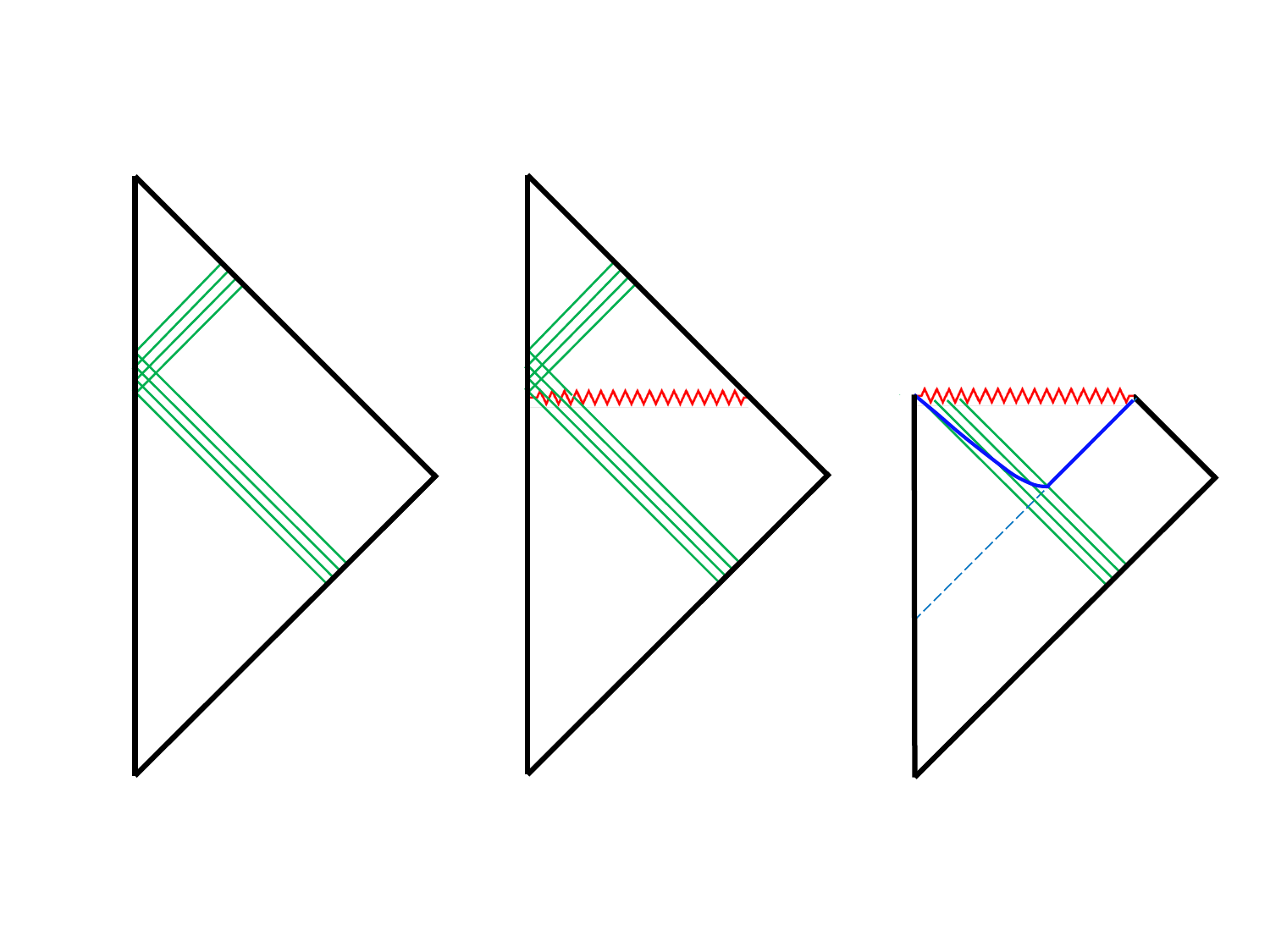}
\end{center}
\vspace{-1cm}
\caption{The first figure displays the half Minkowskian plane $x\geq 0$. The axis is located along the left timelike boundary of the figure at $x=0$. The  two other 
boundaries are past and future null infinity. The lines represent the support of the matter field and its reflection off the axis in accord with Equation (\ref{fp=fm}).
In the second figure, gravity is turned on and a singularity, depicted by the wavy line,  forms as soon as the first strand of matter hits the axis so that the reflected
stream is cut out of the physical spacetime. The physical spacetime is the Vaidya solution depicted in the third figure. The event horizon is along the dotted line.
A 1 parameter family of outer marginally trapped spheres known as a {\em dynamical horizon} (bold line) forms at the left end of the singularity and follows the event horizon after matter collapse. Matter infall is along the unbroken null lines from 
past null infinity to the singularity.
}
\label{figreflectingstreams}
\end{figure}

The stress energy tensor (\ref{tvaidya}) for the solution (\ref{fpm}) takes  the form  of a pair of (infalling and outgoing) spherically symmetric null dust streams.
If there was only an infalling stream, the stress energy would be exactly of the form appropriate to the Vaidya solution.
Note however that:\\
(i) If there is only a single infalling stream with $f$ satisfying the condition of prompt collapse (\ref{pc}), the resulting Vaidya solution
exhibits the following feature: As soon as the first
strand of matter hits the axis a {\em spacelike} singularity forms (see fig \ref{figreflectingstreams}).
\\
(ii) The solution (\ref{fpm}) satisfies reflecting boundary conditions (\ref{reflect}) at the axis. This means that each strand of the null infalling stream
hits the axis and is reflected to an outgoing null stream. Since the singularity of (i) is {\em spacelike}, the outgoing stream is `above' the singularity 
(see fig \ref{figreflectingstreams}). Hence in the {\em physical} spacetime solution we have only the infalling stream.
\\

From (i) and (ii) above, a solution to the classical equations (\ref{gangle})- (\ref{gppmm}) is the Vaidya solution with stress energy tensor
$T_{++} = \frac{1}{4\pi R^2}\frac{(\partial_{+}f)^2}{2}$.
Since the spacetime geometry in this solution is 
flat in a finite neighborhood of the axis, the Vaidya solution satisfies our axis requirements. As shown below, it also satisfies
the initial conditions at $\I^-$.  Hence it is an acceptable solution.

The Vaidya solution is usually presented in Eddington Finkelstein coordinates $(v, R)$ whereas here we use null coordinates. 
The relation between the Eddington Finkelstein (EF) and null coordinates is as follows (the reader may find it easier to follow our argumentation below by 
consulting the Penrose diagram for the Vaidya spacetime depicted in Figure \ref{figreflectingstreams}).

Consider the Vaidya solution for a mass profile $m(v)$ 
at $\I^-$. In EF coordinates  $(v,R)$ the 2-metric is:
\be
{}^{(2)}ds^2= -(1- \frac{2m(v)}{R})(dv)^2  + 2dvdR
\label{efds2}
\ee
with constant $v$ radial lines being null and ingoing. These ingoing light rays originate at $\I^-$ of the Vaidya spacetime where $R\rightarrow \infty$.
These light rays `reflect' off the axis and become  outgoing. Since every   outgoing ray originates at the axis as the reflection of a unique incoming ray, 
we can uniquely label each outgoing ray by the value of $v= v_{axis}:=u$ at this origin point on the axis. Thus constant $u$ light rays are outgoing,
constant $v$ light rays are incoming and every point in Vaidya spacetime is uniquely located as the intersection of a pair of such rays. This implies that $u,v$
are null coordinates. We now show that the identifications  $v\equiv x^+, u \equiv x^-$ hold.

From (\ref{efds2}) it follows that on an outgoing light ray, $R$ changes as a function of $v$ according to
\be
2\frac{dR}{dv} = (1- \frac{2m(v)}{R})  \label{efcc1}.
\ee
Consider the outgoing ray which starts from the axis at $v=v_{axis}$. As discussed above, we set $v_{axis}=u$.
Since $R=0$ at the axis, we may integrate (\ref{efcc1})
to obtain, for the trajectory of this ray:
\be
2R(v,u) = \int_{u}^{v} d{\bar v} (1- \frac{2m({\bar v})}{R({\bar v},u)}).
\label{rvu}
\ee
Let the support of $m(v)$ start on $\I^-$ at $v=v_i$. Then for $v<v_i$, (\ref{rvu}) implies that:
\be
R(v,u) = \frac{v- u}{2}
\label{rvuaxis}
\ee
so that the axis lies at:
\be
v=u
\label{v=u}
\ee
Note that we can rewrite (\ref{rvu}) as:
\be
2R(v,u) = v_i- u + \int_{v_i}^{v} d{\bar v} (1- \frac{2m({\bar v})}{R({\bar v},u)}).
\ee
In this form it is clear that the integrand (and hence the equation) is well defined everywhere except at the $R=0$ singularity. 

Next, note that since $\I^-$ is approached as $R\rightarrow \infty$ along constant $v$, it follows from (\ref{rvu}) that
near $\I^-$:
\be
R(v,u) =  \frac{v- u}{2} + O(\frac{1}{R})
\ee
so that $\I^-$ is approached as $u\rightarrow -\infty$. In this limit, (\ref{rvu}) implies that:
\be 
R(v,u)=  \frac{v- u}{2} + O(\frac{1}{u})
\label{rvuscri}
\ee
Next, note that setting $R= R(v,u)$ in (\ref{efds2}), we have
\be
{}^{(2)}ds^2= -(1- \frac{2m(v)}{R}) (dv)^2 + 2dv( R,_vdv +R,_udu)= [-(1- \frac{2m(v)}{R}) +2 R,_v](dv)^2 + R,_u2dvdu
\ee
which in conjunction with (\ref{efcc1}) implies that in these coordinates the conformal factor $e^{2\rho}$ is given by:
\be
2R,_u =: -e^{2\rho} \label{ef2a}
\ee
From (\ref{rvuscri}) and (\ref{ef2a}) it follows that 
\be
-{2R,_u} =  1 + O(\frac{1}{u^2}) :=e^{2\rho}
\label{iv2}
\ee
Note that from (\ref{rvu}) we have that:
\be
(R,_u),_v= \frac{m(v)}{2R^2} R,_u .
\label{mono}
\ee
Since from (\ref{rvuaxis}) $R,_u <0$ at the axis ,  (\ref{mono}) ensures that $R_u$  remains negative on every outgoing ray, and hence, negative everywhere so that
the identification (\ref{iv2}) is consistent with the postivity of $e^{2\rho}$.

The above analysis shows that $v,u$ are well defined null coordinates for which the axis conditions (\ref{v=u}) and `initial' conditions (\ref{rvuscri}), (\ref{iv2})
are satisfied. Further, the geometry in the vicinity of the axis is flat and hence non-singular. Hence we may identify $v$ with $x^+$ and  $u$ with $x^-$, and
(from the definition of $m(v)$ for Vaidya spacetime)  the mass function  $m(v)$ as:
\be
m(v)= m(x^+)= \int_{x^+_i}^{x^+} d{\bar x^+} \frac{(\partial_{+}f)^2}{2} 
\ee
identical to (\ref{defm}) where the support of $f(x^+)$ in the above equation is between $v_i\equiv x^+_i$ and $x^+_f$.
As a final consistency check, note that from (\ref{mono}), (\ref{ef2a}) we have that
\be
\rho,_+=  \frac{m(x^+)}{2R^2},
\label{efccr2}
\ee
which in conjunction with 
(\ref{efcc1}), (\ref{rvu}) and  (\ref{defm}) suffice to verify that equations (\ref{gangle})-(\ref{gppmm})
are satisfied.

\section{\label{sec4}The semiclassical theory}

\subsection{\label{secfhat}Quantization of the matter field}
Since the matter field satisfies the free wave equation on the fiducial flat spacetime subject to reflecting boundary conditions at $x=0$ it can be
quantized with mode expansion:
\be
{{\hat f}(x^+, x^-)} = 
\int_0^{\infty} dk \frac{\cos kx }{\sqrt{\pi k}}  ({\hat a}(k) e^{-ikt} + {\hat a}^{\dagger}(k) e^{ikt}).
\label{hatf1}
\ee
Defining 
\ba
{\hat f}_{({{+}})}(x^+) &:= & {\int_0^{\infty} dk \frac{1}{\sqrt{4\pi k}} ({\hat a}(k) e^{-ikx^+} + {\hat a}^{\dagger}(k) e^{ikx^+}  )}  \label{fhatp}\\
{\hat f}_{(-)}(x^-)&:= & {\int_0^{\infty} dk \frac{1}{\sqrt{4\pi k}} ({\hat a}(k) e^{-ikx^-} +  {\hat a}^{\dagger}(k) e^{ikx^-} )}  \label{fhatm}
\ea
we may rewrite (\ref{hatf1}) as 
\be
{{\hat f}(x^+, x^-)} ={{\hat f}_{(+)}(x^+)} + {{\hat f}_{(-)}(x^-)}.
\label{hatf=fp-fm}
\ee
Note that the operator valued distribution ${\hat f}_{(+)}$ is the same `operator valued function' of its argument as 
${\hat f}_{(-)}(x^-)$. This is exactly the quantum implementation of the reflecting boundary condition (\ref{fp=fm})

The mode operators  ${\hat a}(k),{\hat a}^{\dagger}(k)$ provide a representation of the classical symplectic structure which follows from the 
matter action (\ref{sfsph}) so that the only non-trivial commutation relations are the standard ones:
\be
[{\hat a}(k),{\hat a}^{\dagger}(l)]= \hbar \delta (k,l),
\label{aadcom}
\ee
which are represented via the standard Fock space representation so that the Hilbert space ${\cal H}_{Fock}$ is the standard Fock space generated by the action of the 
creation operators ${\hat a}^{\dagger}(k)$ on the Fock vacuum.

This quantization may be used to define a test quantum field on the classical Vaidya solution, or to define a quantum field on a general spherically symmetric metric
of the form (\ref{4gsph}) or, as we propose in section \ref{sec6}, to define a quantization of the true degrees of freedom of the combined matter-gravity system.

If we use it to define a 4d spherically symmetric test quantum field (coupled to the 4 metric as in (\ref{smatter}), hence  conformally coupled to the 2-metric $\g2$)
on the Vaidya spacetime, one can put the test scalar field in 
its vacuum state at $\I^-$
and ask for its particle content as experienced by inertial  observers at $\I^+$. 
\footnote{\label{fnfreefall}
From (\ref{aflat}), the $x^{\pm}$ coordinate frame is freely falling at $\I^-$. Hence the Fock vacuum in ${\cal H}_{Fock}$ is the vacuum state for freely
falling observers at $\I^-$.} A straightforward calculation along the lines of 
Hawking's \cite{hawking} leads to the Hawking effect i.e. the state at $\I^+$ exhibits late time  thermal behavior at Hawking temperature $\frac{1}{8\pi M}$.
The calculation is simpler than Hawking's as, due to the 2d conformal coupling, there is no scattering of particles off the spacetime curvature and hence 
no non-trivial grey body factors.

If we use the quantization to define a 4d spherically symmetric quantum field (coupled to the 4 metric as in (\ref{smatter}), hence  conformally coupled to the 2-metric $\g2$)
on a general spherically symmetric metric (\ref{4gsph}), (\ref{g2pm}), we can compute its stress energy tensor expectation value using the results of 
Davies and Fulling \cite{fd}. Note that since the axis serves as a reflecting boundary and since its trajectory is that of a straight line in the
inertial coordinates of the fiducial flat spacetime, the results of Reference \cite{fd} can be directly applied.  

Recall from \cite{fd} that in  the case that 
the initial state
at $x^-\rightarrow -\infty$ is a coherent state in ${\cal H}_{Fock}$ modelled on a classical field $f$, 
the vacuum contribution  to the stress energy expectation value gets augmented by 
the classical stress energy of $f$. Recall (see footnote \ref{fnfreefall}) that the $x^{\pm}$ coordinates are freely falling at $\I^-$ so that 
the initial state is a coherent state as seen by freely falling observers at $\I^-$. Putting all this together we have, from \cite{fd} that the only 
non-trivial components of the 4d stress energy expectation value are given through the expressions
\ba 
8\pi R^2<{\hat T}_{+-}> &= &-\frac{\hbar}{12\pi}\partial_+\partial_-\rho \label{qtpm}\\
8\pi R^2 <{\hat T}_{\pm\pm}> &=& (\partial_{\pm}f)^2 - 
\frac{\hbar}{12\pi} ((\partial_{\pm}\rho)^2 - \partial_{\pm}^2 \rho)  \label{qtppmm}
\ea
The factors of $8\pi R^2$ come from the definition of the 4d stress energy (see (\ref{tvaidya})). In general the expressions in (\ref{qtppmm}) would be 
augmented by functions $t_{\pm}(x^{\pm})$ which are state dependent. Here, these vanish because 
the mode expansion (\ref{hatf1}) is defined
with respect to the $x^{\pm}$ coordinates \cite{fd}.

\subsection{Semiclassical Equations}
The semiclassical Einstein equations find their justification in the large $N$ approximation \cite{hh}. Accordingly we couple $N$ scalar fields
exactly as in (\ref{smatter}), quantize each of them as in the previous section, put one of them in a coherent state modelled on $f$ 
\footnote{ A function $f$ can be uniquely  characterised by its mode coefficients if its Fourier transformation is invertible. In a coherent state, 
the Fourier mode coefficient
of every  positive frequency mode 
is  realised as the eigen value of the corresponding mode operator. Functions $f$ of interest are of compact support at $\I^-$ and satisfy the 
prompt collapse conditon (\ref{pc}) so that the function is not smooth at its initial support (its first derivative is discontinuous). Nevertheless
the function is absolutely integrable and can be chosen to be of bounded variation whereby its Fourier transform is invertible (see Appendix \ref{secaft} for details).}
and the rest in their
vacuum states at $\I^-$. 
From (\ref{qtpm}), (\ref{qtppmm}) and (\ref{gangle})-(\ref{gppmm}), it then follows that the semiclassical Einstein equations, $G_{ab}= 8\pi \bra T_{ab}\ket$ 
take the form:
\ba
-\frac{e^{2\rho}}{4}G_{{\hat \Omega} {\hat \Omega}} &= & \partial_+\partial_- \rho + \frac{1}{R}\partial_+\partial_-R=0 \label{gangles}\\
R^2\;G_{+-}&=& 2R\partial_+\partial_-R + 2 \partial_+R\partial_-R + \frac{1}{2} e^{2\rho}=-\frac{N\hbar}{12\pi}\partial_+\partial_-\rho
\label{gpms}\\
R^2\;G_{\pm \pm}&=&  R^2[-\frac{2}{R}( \partial_{\pm}^2R- 2\partial_{\pm}\rho \partial_{\pm}R)]= (\partial_{\pm}f)^2 - 
\frac{N\hbar}{12\pi} ((\partial_{\pm}\rho)^2 - \partial_{\pm}^2 \rho)
\label{gppmms}
\ea

\subsection{\label{secsing}Semiclassical Singularity }

We are interested in semiclassical solutions in which the axis is located at $x=0$, the axial geometry is non-singular and for which the 
asymptotic conditions (\ref{aflat}) hold.
Note that when $\rho=0$ the vacuum fluctuation contribution to the stress energy expectation value vanishes. Thus, when $f$ vanishes, 
classical flat spacetime (with $e^{2\rho}=1, R= \frac{x^+-x^-}{2}$) remains a solution. Hence for $x^+ <x^+_i$ we set the spacetime
to be flat with 
\be
e^{2\rho}=1, \;\;\;R= \frac{x^+-x^-}{2}
\label{initialmm}
\ee

Next, note that we can eliminate $\partial_+\partial_- \rho$ between the first two eqns to obtain:
\be
\frac{1}{R}\partial_+\partial_-R = - \frac{\partial_+R\partial_-R + \frac{1}{4} e^{2\rho}}{R^2- {\frac{N\hbar}{24\pi}}}
\label{evol1}
\ee
Following Lowe \cite{lowe} and Parentani and Piran \cite{pp}, we can look upon (\ref{evol1}), (\ref{gangle}) as evolution equations for initial data
(i) on the null line $x^+= x^{+}_i$ and (ii) on  $\I^-$ for $x\geq x^+_i$ .  For (i) , the initial data is  given by (\ref{initialmm}). 
For (ii), 
the  matter data is subject to (\ref{pc}) and  the  gravitational data corresponds to that for the  
Vaidya solution
with $m(x^+)$ given by (\ref{defm}) and $R, \rho$ obtained by integrating  (\ref{efcc1}) along $\I^-$ and  then using (\ref{ef2a}).
More in detail, at $x^+=x^+_i$, we have data near $\I^-$ of the form (\ref{initialmm}). Equation (\ref{efcc1}) can be integrated along $\I^-$ with this initial data
for $R$ to obtain $R$ along $\I^-$ for $x\geq x^+$ and $e^{2\rho}$ can then be obtained near $\I^-$ from (\ref{ef2a}).
It can then be shown that the evolution equations can be solved uniquely for $R, \rho$ in the region $x^+ \geq x^+_i$ as long as the evolution equations themselves
are well defined. 

From a numerical evolution point of view \cite{lowe} one can see this as follows.  Along $x^+=x^+_i$, equation (\ref{evol1}) can be viewed as a first order differential equation 
for $R,_{+}$ on $x^{+}=x^+_i$ with `inital' value 
for $R,_{+}$ specified on $\I^-$ and known coefficients from (\ref{initialmm}). The solution $R,_+ (x^+=x^+_i, x^-)$ together with the initial value for $\rho,_+$
on $\I^-$ can be used to solve (\ref{gangle}) for $\rho,_+$ on the line $x^{+}=x^+_i$. From this one has data $\rho,R$ for the next $x^+$=constant=$x^+_i+ \e$ line on the 
numerical grid and the procedure can be  iterated so as to eventually cover all of $x^+>x^+_i$.

We now argue 
that for generic matter data the evolution equations break down at 
$R^2= {\frac{N\hbar}{24\pi}}$ and a curvature singularity develops.
In this regard note that the denominator of the right hand side of (\ref{evol1}) blows up at $R^2= {\frac{N\hbar}{24\pi}}$.
If the numerator is non-zero at this value of $R$ the left hand side blows up and through (\ref{gangle}) so does $\partial_+\partial_-\rho$.
Since (as can be easily checked), $\2R = 8 e^{-2\rho}\partial_+\partial_-\rho $ we expect a 2-curvature singularity at this value of $R^2$.
If the numerator vanishes at some $x^+=a^+ >x^+_i, x^-= a^-$ where $R^2(a^+,a^-)= {\frac{N\hbar}{24\pi}}$, 
one can slightly change the initial data for $f$ on $\I^-$, thereby change the function $R$ along
$\I^-$ (for $x^+>x^+_i$) and hence the `initial' data $R,_+$ for (\ref{evol1}) at $x^+=a^+$ on $\I^-$. This would generically result in a change of the 
numerator away from zero. Thus one expects that for generic matter data there is a singularity at $R^2= {\frac{N\hbar}{24\pi}}$.

Thus the `initial point' of the Vaidya singularity of the classical theory moves `downwards' along the initial matter infall line $x^+=x^+_i$  away from the axis 
where $R=0$ 
to $R= \sqrt{ {\frac{N\hbar}{24\pi}} }$ (see Figure \ref{figsemiclassical}).

\subsection{\label{secah}Outer Marginally Trapped Surfaces}

One possible quasilocal characterization of a black hole is the existence of outer marginally trapped surfaces (OMTS's) \cite{hayward,aadh}. 
In this section we
analyse the behavior of spherically symmetric OMTS's in the context of the system studied in this work. To this end, fix an $R=$constant 2 sphere.  Let the 
expansions $\theta_+$ and $\theta_-$ denote the expansions of outward and inward future pointing radial null congruences at this sphere. The sphere is defined to be an 
OMTS if $\theta_+=0, \theta_-<0$. A straightforward calculation yields:
\be
\theta_{\pm}= 2 e^{-2\rho} \frac{\partial_{\pm}R}{R}
\label{exppm}
\ee
Since the physical spacetimes considered in this work are flat near the axis, such OMTS's can only form in these solutions away from the axis where $R>0$.
Hence (assuming we are away from singularities), the conditions for an OMTS to form are:
\ba
\partial_+R &=& 0 \label{pr=0}\\
\partial_- R &<&0  \label{mr<0}
\ea
While an OMTS is a quasilocal characterization of a black hole at an `instant of time' and hence a 2-sphere, the quasilocal analog of the 3d event horizon
is a 1 parameter family of OMTS's which form a tube which we call an Outer Marginally Trapped Tube (OMTT). 
The shape of a spherically symmetric OMTT 
(i.e. a tube foliated by spherically symmetric OMTS's) can be studied as follows.
Since (\ref{pr=0}) holds, the  normal $n_a$ to the OMTT is:
\ba
(n_+, n_-) = (\partial_+^2R, \partial_-\partial_+R) 
=  (-4\pi R \bra T_{++}\ket,  -\frac{Re^{2\rho}}{4(R^2- \frac{\hbar N}{24\pi})}) \label{norm2}\\
\Rightarrow n^an_a = -4e^{-2\rho}4\pi R \bra T_{++}\ket \frac{Re^{2\rho}}{4(R^2- \frac{\hbar N}{24\pi})}
\label{n2}
\ea
where we have used the `++' equation in (\ref{gppmms}) with (\ref{pr=0}) to calculate $n_+$ and  (\ref{evol1}) with (\ref{pr=0}) to compute $n_-$ in (\ref{norm2}).
From (\ref{n2}), $n_an^a$ is  timelike, spacelike  or null  if $\bra T_{++}\ket$ is, respectively positive, negative or vanishing so that OMTT
is, respectively, spacelike, timelike or null.

Following \cite{susskind}, we coordinatize the trajectory of the spherically symmetric OMTT by $x^+$ and study how $x^-$ changes with $x^+$ along this trajectory.
Since $\partial_+R$ vanishes along this trajectory we have that:
\ba
\frac{d\partial_+ R}{dx^+} &= &\partial_+^2R + \frac{dx^-}{dx^+} \partial_-\partial_+R = 0 \nonumber\\
\Rightarrow \frac{dx^-}{dx^+} &=& -\frac{ \partial_+^2R}{\partial_-\partial_+R }= -(16\pi e^{-2\rho}  (R^2- \frac{\hbar N}{24\pi})) \bra T_{++}\ket \label{shape1}
\ea
where we have used (\ref{norm2}). Equation (\ref{shape1}) leads us to the same correlation between positivity properties of the stress energy and the 
spacelike, timelike or null nature of the OMTT as above.

Next, note that on the OMTT:
\ba
\frac{dR}{dx^+} &=& \partial_+R + \frac{dx^-}{dx^+} \partial_-R \nonumber\\
&=&  (16\pi e^{-2\rho}  (R^2- \frac{\hbar N}{24\pi}))  (-\partial_-R)  \bra T_{++}\ket \label{rprop}
\ea
where we have used (\ref{pr=0}) together with (\ref{shape1}). From (\ref{mr<0}) it follows that $R$ (and hence the area of the OMTS cross section of the
OMTT) increases, decreases or is unchanged if, respectively, $\bra T_{++}\ket$  is positive, negative or null.

The above set of results correspond to those of \cite{aadh}  restricted to  the 
simple spherically symmetric setting of our work.
\footnote{We note here that the notion of an OMTT corresponds to that of a {\em Trapping Dynamical Horizon}. The notions of Trapping and Anti-trapping Dynamical Horizons
are described in \cite{aapersp}
and constitute a refinement for semiclassical purposes of the notion of Dynamical Horizons introduced in \cite{aadh}, the latter notion being adapted for anticipated
applications in classical (numerical) gravity. 
}
Let us apply them to the following physical scenario. For a large black hole, we expect a low Hawking temperature and low rate of thermal emmission.
As we shall see in section \ref{sec5}, the Hawking emmission
in a QFT in CS calculation goes as $N\hbar M^{-2}$ at $\I^+$. Consequently, for our purposes here the condition $M^2 >> N\hbar $ characterizes a `large' black hole.
\footnote{Restoring factors of $G$, this condition reads, in units in which $c=1$, as $(GM)^2 >>N\hbar G$.}

Let us assume that the collapse lasts for a small duration (i.e. $x^+_f -x^+_i << GM$) during which classical infall dominates quantum back reaction
at large $R$ (including at $R\sim M$). Once the collapse is over, we expect the black hole to start radiating slowly. We can estimate the local rate of mass loss
due to this radiation by assuming the geometry at this epoch is well approximated by the classical Vaidya geometry. More precisely, let us assume that 
the quantum radiation starts along the line $x^+=x^+_f$ at the 2-sphere  at which the event horizon $R=2M$ intersects this line. Since this 2 sphere is an OMTS
in the Vaidya spacetime, within our approximation we may apply (\ref{rprop}) to estimate the rate of change of area of this OMTS with the right hand side 
calculated using the Vaidya geometry:
\ba 
\frac{dR}{dx^+} &=& (16\pi e^{-2\rho}  (R^2- \frac{\hbar N}{24\pi}))  (-\partial_-R)  \bra T_{++}\ket \\
&\approx &  (-2\partial_-R e^{-2\rho}) R^2 8\pi \bra T_{++}\ket \\
&\approx &  -\frac{N\hbar}{12\pi} ((\partial_{\pm}\rho)^2 - \partial_{\pm}^2 \rho) \label{ehflux}
\ea
where in the second line we used  the large black hole approximation $(M^2 >> N\hbar $ and in the third we used the property (\ref{ef2a}) of the Vaidya spacetime
together with equation (\ref{gppmms}). Using (\ref{efccr2}) with $R=2M$, we have  $\partial_+\rho = \frac{1}{8M}$ and, using (\ref{efccr2}) together with 
(\ref{efcc1}) we have $(\partial_+)^2 \rho =0$. Putting this in (\ref{ehflux}) and setting $R=2M$ on the left hand side, we obtain
\be
\frac{dM}{dx^+} \approx -\frac{N\hbar}{24\pi} \frac{1}{64M^2}
\label{ehflux1}
\ee
Remarkably this agrees with rate of mass loss obtained at $\I^+$ (see (\ref{fluxv}) of section \ref{sec5}).
This agreement of quasilocal mass loss with that at $\I^+$ for large black holes also seems to happen for the case of CGHS black holes \cite{susskind}.
While this agreement is expected on physical grounds, we do not have a deeper technical understanding for this agreement; for example, a proof
of agreement based on stress energy conservation together with the large black hole approximation would constitute such an understanding.

\subsection{\label{secpplowe}The semiclassical spacetime solution: folding in results from prior numerics}

While the semiclassical equations do not seem amenable to analytical solution, the particular semiclassical
solution of interest with flat geometry in a finite region around the symmetry axis { is} amenable to {\em numerical} solution along the lines reviewed in 
section \ref{secsing}. While we advocate
a careful numerical study along the lines of \cite{franz}, there are two prior numerical works by Lowe \cite{lowe} and by Parentani and Piran \cite{pp}
which are of relevance. While these beautiful  works are not cognizant of key aspects of  the coherent picture developed in this work, the semiclassical equations they solve
are 
practically 
the same as those in this work and they provide a key complimentary resource to our work here.

The work by Lowe \cite{lowe} uses exactly the same action (\ref{ssph}) (modulo some overall numerical factors) and hence obtains the same 
semiclassical equations (modulo some numerical factors). Since the importance of the axis and the axis reflecting boundary conditions for the 
matter field is not realised, the state dependent functions $t_{\pm}(x^{\pm})$ (see the end of section \ref{secfhat}) are not pre-specified 
but chosen in accord with the physical situation which is modelled.
The `classical' component of  matter is chosen to be a shock wave 
with a Dirac delta stress energy along this infall line (from our point of view the Dirac delta function ensures that the prompt collapse condition (\ref{pc}) is 
satisfied). 
Along the infall line the data for $R,\rho$ is chosen as $R=\frac{x^+_i-x^-}{2}, \rho=0$.  
Initial conditions at $\I^-$ beyond the point of matter infall are specified which correspond to the asymptotic behavior of the Schwarzschild solution.
These suffice for well defined numerical evolution as described
 in section \ref{secsing}. 

One difference with our work here is that 
these conditions by virtue of the presence of logarithmic terms in metric fall offs at $\I^-$ \cite{lowe}, do not agree with the conditions (\ref{aflat}). 
We believe, contrary to the implicit  assertion in Reference \cite{lowe}, that  a continuation of  the  data of \cite{lowe} 
on $x^+=x^+_i$ to flat spacetime data $R=\frac{x^+_-x^-}{2}, \rho=0$ for $x^+<x^+_i$, 
is in contradiction with the behavior of the Vaidya solution near $\I^-$. While it  would be good to clarify whether such a continuation is consistent 
with the Einstein equations near $\I^-$, this is beyond the scope of our paper. Notwithstanding this, we shall assume that the physics which emerges from
the numerical results of \cite{lowe} is robust enough that it applies to the system studied in this work.


Lowe \cite{lowe} notes the existence of a {\em spacelike} semiclassical singularity and the emanation of an OMTT at the infall line.
\footnote{Following the approach of Reference \cite{susskind}, we have integrated the evolution equation (\ref{evol1}) just beyond the infall line of a shock
wave, used junction conditions consistent with our asymptotic behavior (\ref{aflat}), 
and verified the existence of the semiclassical singularity at $R^2= \frac{\hbar N}{24}$ and the emanation of an OMTT on the infall line when 
$R^2= 4M^2+ \frac{\hbar N}{24}$.}
Since the classical matter is a shock wave with no extended support, quantum backreaction starts immediately and the OMTT is timelike. The OMTT and the singularity meet
away from $\I^+$ in the interior of the space time. The outgoing future pointing radial null rays starting at this intersection form a Cauchy horizon. 
There is no evidence of a `thunderbolt' along this
`last' set of null rays to $\I^+$. This `outer' Cauchy horizon is in addition to the 
`inner' Cauchy horizon which forms along the infall line beyond  $R^2= \frac{\hbar N}{24}$. 

Parentani and Piran \cite{pp} define the semiclassical equations without recourse to action based arguments by positing the stress energy tensor to be the sum
of a classical part and a quantum back reaction part. The former is posited to be of the null dust type infall appropriate to Vaidya. The profile of the dust is
chosen to be a Gaussian but in the numerics we are unable to discern if its tail is cut off and if so whether, effectively, the prompt collapse condition (\ref{pc}) 
is satisfied. While the work explicitly recognizes the existence of an axis, its import for the reflecting boundary conditions in the quantization of the 
scalar field (see section \ref{secfhat}) is not recognized. The fact that the classical solution is Vaidya and that for a Gaussian profile 
which is not one of prompt collapse,
the classical singularity structure is complicated \cite{bengtsson} is not appreciated.
\footnote{This brings up the extremely interesting question: do back reaction effects cause the complicated (locally/globally) naked singularity structure of the 
classical solution for non-prompt collapse to simplify?}
The quantum contribution to the stress tensor is chosen to be of  exactly 
the form in (\ref{gpms})-(\ref{gppmms}) without the realization that it could arise naturally through quantization of an appropriately chosen classical 
scalar field as shown in this work. 

The solution chosen is, by virtue of ignoring the tail contributions of the Gaussian, in practice flat in a finite neighborhood of the axis so that  the 
dynamical and constraint equations and set up for numerical evolution are exactly the same as Lowe. Since the set up is numerical, initial conditions are
at large but finite $x^-=x^-_I$ rather than at $\I^-$. A coordinate choice of $x^+$ which agrees with ours for $x^+<x^+_i$ but differs from ours elsewhere
is made. This choice depends on $x^-_I$ and as $x^-_I\rightarrow -\infty$  approaches ours. We shall assume that the basic physics is robust 
with regard to the difference in these choices.

Parentani and Piran note the existence of a spacelike  semiclassical singularity at $R^2= \frac{\hbar N}{24}$ and a OMTT which is spacelike 
as long as classical matter infall dominates after which it turns timelike and meets the singularity away from $\I^+$. Similar to \cite{lowe} a Cauchy horizon then forms.
Interestingly, the quantum flux at  $\I^+$ starts out as thermal flux at temperature inversely proportional to the initial mass, its mass dependence being $\sim M^{-2}$
as expected. However at late stages of evaporation, 
near the  intersection of $\I^+$   
with the Cauchy horizon,  
where the Bondi 
mass gets small, the flux turns around to a less divergent function of this small mass. We take this as evidence for lack of a thunderbolt.

\begin{figure}
\begin{center}
\includegraphics[width= 0.4\textwidth]{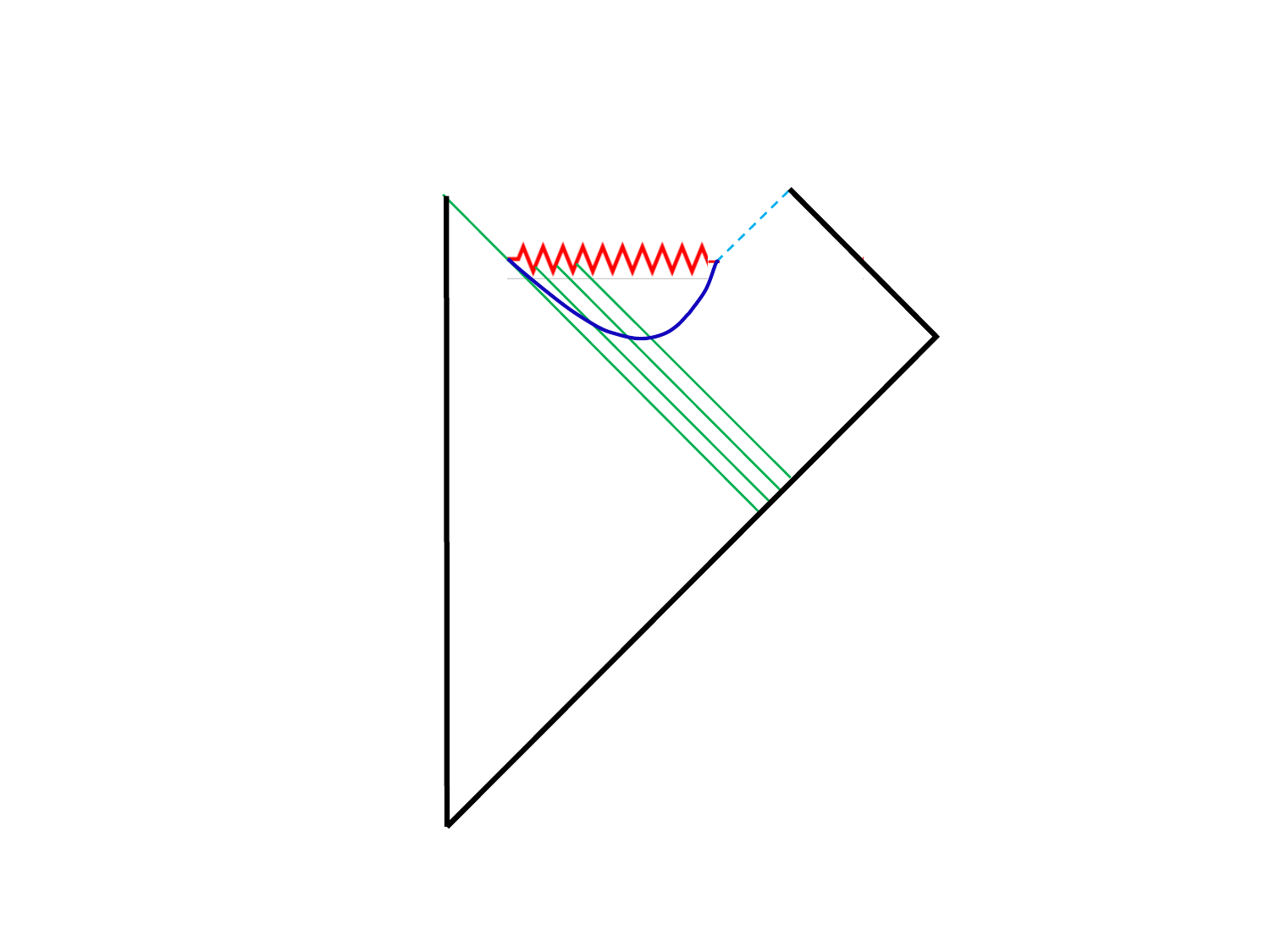}
\end{center}
\vspace{-1cm}
\caption{The figure depicts the proposed semiclassical spacetime solution. Towards the left of the initial  matter infall line, the spacetime is flat.
The singularity, depicted by a wavy line, starts along this infall line  and is located away from the axis at  $R =\sqrt{\frac{N\hbar}{24}}$. A spacelike outer 
marginally  trapped tube (i.e. spacelike dynamical horizon)
horizon
is born at the left end of the singularity and grows as long as the classical stress energy dominates the quantum contribution. Once the back reaction dominates 
the effect of the classical matter stress energy, the spacelike dynamical horizon becomes timelike i.e. it turns upwards into
a timelike marginally trapped tube which meets the singularity. A Cauchy horizon forms along the last rays which originate  at this meeting point and travel to $\I^+$.
There is also a Cauchy horizon between the left end of the singularity and the axis.}
\label{figsemiclassical}
\end{figure}

Putting together (i) the  analytical work of sections \ref{secsing},  \ref{secah}, (ii) the  physical intuition that the  initial part of spacetime 
is dominated by classical collapse 
followed by quantum radiation and (iii) the beautiful numerical work of References \cite{lowe,pp}, we propose the Penrose diagram in Fig \ref{figsemiclassical} as a description of the semiclassical 
spacetime geometry.

\section{\label{sec5} A balance law at $\I^+$}

In this section we show that the semiclassical equations at $\I^+$ imply  a balance law which relates the rate of decrease of a {\em back reaction corrected} Bondi 
mass to a {\em manifestly positive}
back reaction corrected Bondi flux at $\I^+$. The derivation of the balance law rests on two physically reasonable assumptions which we detail in the next paragraph.
These assumptions together with the semiclassical equations relate the rate of change of Bondi mass to the stress energy expectation value at $\I^+$.
The stress energy expectation value is not manifestly positive. However, following  identical considerations in the analysis of the evaporation of 
2d dilatonic black holes in Reference \cite{atv}, a back reaction corrected and manifestly positive Bondi flux can be identified which drives the rate of decrease of
a back reaction corrected Bondi mass. We proceed to the detailed derivation.

We make two assumptions regarding the asymptotic behavior of the metric at $\I^+$:\\

\noindent A1. We expect  that at early times at $\I^+$ back reaction effects have not built up and that the classical {\em ingoing} Vaidya solution discussed in section 
\ref{sec3} is a 
good approximation to the  spacetime geometry. \\

\noindent A2. We expect that eventually back reaction effects build up and produce a non-trivial stress energy flux at  $\I^+$. Since the system is spherical symmetric
we assume that this situation can be modelled by an {\em outgoing} Vaidya metric all along  $\I^+$. Thus the metric near $\I^+$ is assumed to take the form:
\be
{}^{(2)}ds^2= -(1- \frac{2m_B(\u)}{R})(d{\bar u})^2  - 2d\u dR +O(\frac{1}{R^2}).
\label{efout}
\ee 
Here  $R\rightarrow \infty$, $\u$ is an  Eddington-Finkelstein null coordinate and the subscript $B$ on the outgoing mass indicates that this mass
is the Bondi Mass.

From A1, at early times, $\I^+$ is located at $x^+=\infty$. Since $\I^+$ is null, we shall assume that at all times, it is located at $x^+= \infty$.
Since $\u$ is an outgoing null coordinate,
the 2-metric (\ref{efout}) near $\I^+$ can be expressed in conformally flat form in the coordinates $x^+, \u$ as:
\be
{}^{(2)}ds^2=  -e^{2\brho} dx^+ d\u
\label{ccbar}
\ee
Similar to (\ref{efccr2}), to leading order in $\frac{1}{R}$, it follows that
\be
\frac{\partial}{\partial \u}\brho (x^+, \u)  = O(\frac{1}{R^2})\;\;\;\;\; \frac{\partial^2}{\partial \u^2}\brho (x^+, \u )= O(\frac{1}{R^2})
\label{neglect}
\ee
Since $\u, x^-$ are both  outgoing null coordinates, the coordinate $\u$ is a function
only of $x^-$ and not $x^+$. 
Using this fact together with the `$--$' constraint (\ref{gppmms}), the asymptotic form (\ref{ccbar}) and the behavior of the conformal factor
(\ref{neglect}), it is straightforward to show that at $\I^+$ in the limit $R\rightarrow \infty$:
\ba
-\frac{1}{2}R^2G_{{\bar u}{\bar u}} &= & \frac{dm_{B}}{d{\bar u}} \label{balance1}\\
&=& -4\pi R^2 \bra {\hat T}_{\bar u\bar u}\ket \\
&=& 
-{\frac{1}{2}(\partial_{\bar u} f)^2} + 
\frac{N\hbar}{48\pi}(\frac{1}{ {\bar u}^{\prime} })^2[ \frac{3}{2}(\frac{{\bar u}^{\prime \prime}}{ {\bar u}^{\prime} })^2 
- \frac{ {\bar u}^{\prime\prime\prime}   }{  {\bar u}^{\prime}  }] \label{balance2}
\ea
where each  `$\prime$' superscript signifies a derivative with respect to $x^-$ so that, for e.g.  $\u^{\prime}:= \frac{d\u}{dx^{-}}$.
Thus we have derived a balance law relating the change of Bondi mass (with respect to the asymptotic translation in $\u$ along $\I^+$) to 
the energy flux at $\I^+$:
\be
\frac{dm_{B}}{d{\bar u}} = -{\frac{1}{2}(\partial_{\bar u} f)^2} + 
\frac{N\hbar}{48\pi}(\frac{1}{ {\bar u}^{\prime} })^2[ \frac{3}{2}(\frac{{\bar u}^{\prime \prime}}{ {\bar u}^{\prime} })^2 
- \frac{ {\bar u}^{\prime\prime\prime}   }{  {\bar u}^{\prime}  }] := -{\cal F}
\label{balance3}
\ee
The energy flux ${\cal F}$ has a `classical' part  ${\cal F}^{classical}$ 
corresponding to the first  term on the right hand side of (\ref{balance3})
and a quantum backreaction part ${\cal F}^{quantum}$ corresponding to the rest of the right hand side of (\ref{balance3}):
\footnote{Both these contributions arise from the stress energy {\em expectation} value and hence are, ultimately, quantum in origin.
The first term on the right hand side of (\ref{balance3}) arises by virtue of our choice of initial state as a coherent state patterned on classical data $f$ and depends exclusively 
on $f$ with no dependence on $\hbar$, whereas the rest of the expression has an explicit $\hbar$ dependence hence the choice of nomenclature.}
\ba
{\cal F} &=& {\cal F}^{classical} + {\cal F}^{quantum} \label{fluxtotal} \\
{\cal F}^{classical} &=& {\frac{1}{2}(\partial_{\bar u} f)^2}\label{fluxclass}\\
{\cal F}^{quantum} &=&
-\frac{N\hbar}{48\pi}(\frac{1}{ {\bar u}^{\prime} })^2[ \frac{3}{2}(\frac{{\bar u}^{\prime \prime}}{ {\bar u}^{\prime} })^2 
+ \frac{ {\bar u}^{\prime\prime\prime}   }{  {\bar u}^{\prime}  }] 
\label{fluxquantum}
\ea
While the classical  piece is explicitly positive definite, this property does not hold for the quantum piece.
However, following \cite{atv}, we can rewrite this quantum part as:
\ba
{\cal F}^{quantum} &=&
-\frac{N\hbar}{48\pi}(\frac{1}{ {\bar u}^{\prime} })^2[ \frac{3}{2}(\frac{{\bar u}^{\prime \prime}}{ {\bar u}^{\prime} })^2 
- \frac{ {\bar u}^{\prime\prime\prime}   }{  {\bar u}^{\prime}  }] \\
&=& [\frac{d}{d{\bar u}} {\frac{N\hbar}{48\pi}(\frac{ {\bar u}^{\prime \prime}}{( {\bar u}^{\prime})^2}}]
+\frac{N\hbar}{96\pi}\frac{({\bar u}^{\prime \prime})^2}{( {\bar u}^{\prime})^4 }.
\label{balance4}
\ea
Using (\ref{balance4}) we may rewrite (\ref{balance3}) as:
\be
\frac{d}{d{\bar u}}[m_{B} +  {\frac{N\hbar}{48\pi}(\frac{ {\bar u}^{\prime \prime}}{( {\bar u}^{\prime})^2}}]
=-{\frac{1}{2}(\partial_{\bar u} f)^2} 
-\frac{N\hbar}{96\pi}\frac{({\bar u}^{\prime \prime})^2}{( {\bar u}^{\prime})^4 }.
\label{balance5}
\ee
The right hand side of (\ref{balance5}) is now explicitly negative definite. Equation (\ref{balance5}) suggests that we identify the term in square 
brackets on its left hand side as a back reaction corrected Bondi mass $m_{B,corrected}$,
\be
m_{B,corrected}:=m_{B} +  {\frac{N\hbar}{48\pi}\frac{ {\bar u}^{\prime \prime}}{( {\bar u}^{\prime})^2}},
\label{mcorrect}
\ee
which {\em decreases} in response to the outgoing {\em positive definite} back reaction corrected flux ${\cal F}_{corrected}$ received at $\I^+$ :
\be
{\cal F}_{corrected} := {\frac{1}{2}(\partial_{\bar u} f)^2} 
+\frac{N\hbar}{96\pi}\frac{({\bar u}^{\prime \prime})^2}{( {\bar u}^{\prime})^4 }.
\label{fluxcorrect}
\ee
The form of the back reaction corrected balance law (\ref{balance5}) suggests that black hole evaporation ceases when the corrected Bondi mass $m_{B,corrected}$
(\ref{mcorrect}) is exhausted, at which point the corrected flux ${\cal F}_{corrected}$ (\ref{fluxcorrect}) also vanishes. For ${\cal F}_{corrected}$ to vanish both its `classical' 
and quantum contributions must vanish seperately since both are positive definite. In particular the quantum contribution 
${\cal F}^{quantum}_{corrected}$
must vanish so that:
\be
{\cal F}^{quantum}_{corrected}:= \frac{N\hbar}{96\pi}\frac{({\bar u}^{\prime \prime})^2}{( {\bar u}^{\prime})^4 } =0
\label{qcflux=0}
\ee
  Assuming that this happens smoothly, it must be the case that:
\be
{\bar u}^{\prime \prime}=0 
\ee
which implies that $\u$ is a {\em linear} function of $x^-$:
\be
\u= ax^- +b
\label{ilong}
\ee
for some constants $a,b$. Since $\u$ is a future pointing null coordinate, we have that $\u^{\prime}>0$ and hence, $a>0$.
This implies that as $\u\rightarrow \infty$, $x^-\rightarrow \infty$ which means that 
$\I^+$ of the physical spacetime is `as long' as that of the fiducial Minkowski spacetime. 
Note that in contrast the $\I^+$ of Vaidya ends at $x^-=x^-_H$ where $x^-_H$ is the (finite) value of $x^-$ at the horizon.
In this sense $\I^+$ of the physical spacetime is `quantum' extended beyond its classical counterpart.
This is the main conclusion of this section.
We discuss its possible implications in the next section where we also discuss the origin of the `classical' contribution to ${\cal F}$.

Before doing so, we note that it is possible to calculate the flux ${\cal F}$ (\ref{fluxtotal})  at $\I^+$ of the Vaidya spacetime with $\rho$ corresponding to that of 
the Vaidya solution. Recall from section \ref{sec3} that in the Vaidya solution the outgoing classical flux is absent.
As shown in the Appendix \ref{secaflux}, the quantum flux ${\cal F}^{quantum}$ evaluates at late times on $\I^+$ of the Vaidya spactime to:
\be
{\cal F}^{quantum}= \frac{N\hbar}{24\pi}(\frac{1}{64M^2})
\label{fluxv}
\ee
We can also calculate the corrected quantum flux ${\cal F}^{quantum}_{corrected}$
\be
{\cal F}^{quantum}_{corrected}:= \frac{N\hbar}{96\pi}\frac{({\bar u}^{\prime \prime})^2}{( {\bar u}^{\prime})^4 }
\ee
and as shown in Appendix \ref{secaflux} this agrees with ${\cal F}^{quantum}$ at late times i.e. at late times on $\I^+$, 
\be
{\cal F}^{quantum}_{corrected}= {\cal F}^{quantum}= \frac{N\hbar}{24\pi}(\frac{1}{64M^2})
\label{fqcv}
\ee
Equation (\ref{fqcv}) corresponds to the thermal Hawking flux measured at $\I^+$ in the quantum field theory on curved spacetime approximation.

\section{\label{sec6} Speculations on the deep quantum behavior of the system}

We propose that the true degrees of freedom of the  system (\ref{ssph})  are those of the scalar field and that 
 the gravitational degrees of freedom can be solved for in terms of specified matter data (classically) or when the 
quantum state  of matter is specified (semiclassically and at the deep quantum gravity level). This proposal is supported by the fact that in the classical
theory if we set the matter field to vanish, flat spacetime is the unique classical solution to equations (\ref{gangle})- (\ref{gppmm}) subject to asymptotic flatness at 
past null infinity  (\ref{aflat}) as well as the condition that the   axis of symmetry exist and be located at (\ref{axislocation}).  
\footnote{We have checked this explicitly. The result may be interpreted as an implementation of Birkhoff's theorem.}
Clearly, the proposal would be on a firmer footing if for the classical and semiclassical equations, we could prescribe precise boundary conditions on the geometry 
variables $\rho,R$
at the axis together with the initial conditions
(\ref{aflat}) such that a specification of matter data at $\I^-$ subject to reflecting boundary conditions at the axis (as discussed in  section \ref{secmkin}),
results in a unique solution.  While a complete treatment is 
beyond  the scope of this paper, in anticipation of future work towards such a treatment, we initiate an analysis of possible boundary conditions at the axis 
in Appendix \ref{secaxis} and comment
on the complications which arise due to its timelike nature.

Notwithstanding the remarks above, let us go ahead and assume that the true degrees of freedom at the classical level are those of the classical scalar field data 
at $\I^-$ and that, correspondingly, the true quantum degrees of freedom of the gravity-matter system are those of the quantum scalar field. This implies that the 
Hilbert space for the quantum gravity-matter system is the Fock space ${\cal H}_{Fock}$ constructed in section \ref{secfhat} and that the natural arena for 
these degrees of freedom is the Minkowkskian half plane $x\geq 0$. This assumption is supported by the considerations of section \ref{sec5} wherein we
argued that the physical $\I^+$ was as long as the fiducial Minkowskian $\I^+$.

More in detail the starting point for this argument in section \ref{sec5} is an assumed  validity of the semiclassical equations at $\I^+$. 
These equations via the arguments of \cite{hh} 
relate the Einstein tensor of the expectation value of the metric to the expectation value of the stress energy tensor and are assumed
to hold when the quantum fluctuations of the geometry are negligible.
Hence, while the semiclassical equations are {\em not}  expected to hold near the singularity where geometry fluctuations are expected to be 
significant, it seems reasonable to assume that they do hold near $\I^+$. If they do hold near $\I^+$ (assuming, of course that the expectation value
geometry is asymptotically flat), then the reasonable assumptions of section \ref{sec5} lead to the conclusion of a quantum extended $\I^+$ which is as long as the fiducial Minkowskian 
$\I^+$; this conclusion  is 
supportive of the idea that the correct physical arena is the half Minkowskian plane.

Note also that the proposed true degrees of freedom, namely those of the quantum scalar field, propagate on the fiducial flat spacetime by virtue of their 2d conformal coupling.
Hence these degrees of freedom admit well defined propagation through the semiclassically singular region. 
The infalling quantum scalar field is reflected off the axis transmuting thereby to the outgoing scalar field which registers on the quantum extension of $\I^+$.
This is the origin of the `classical' contribution (\ref{fluxclass}) to the asymptotic flux of section \ref{sec5}.

Since quantum evolution of the true (matter) degrees of freedom of the system is well defined even at classically or semiclassically singular regions,
one might hope that it is possible to define the action of operator correspondents of gravitational variables in these regions as well.
In this sense one may hope that the deep quantum theory resolves the singularities of the classical/semiclassical theory.

\begin{figure}
\begin{center}
\includegraphics[width= 0.5\textwidth]{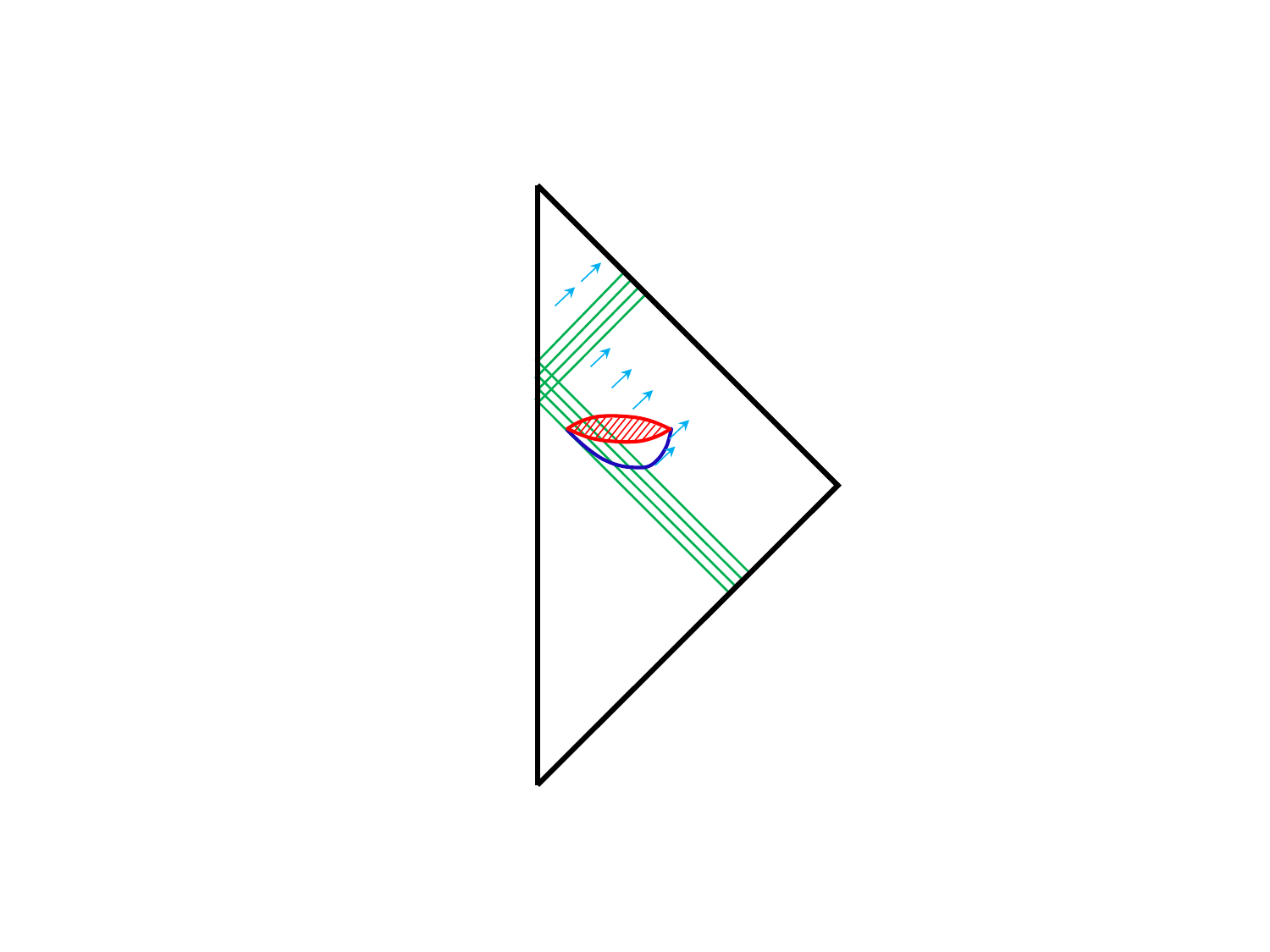}
\end{center}
\vspace{-1mm}
\caption{The figure depicts the quantum extended spacetime manifold which coincides with the half Minkowskian plane. The oval striped region indicates the 
semiclassically singular region which is resolved at the deep quantum level. In this part of the manifold , spacetime geometry fluctuations are large and there is 
no clear notion of causality. A vast region of spacetime opens up beyond this hitherto singular region. The classical matter field, depicted by unbroken lines, passes through this region, is reflected
off the axis and arrives at the extend $\I^+$. The arrows correspond to outgoing quantum stress energy.}
\label{figab}
\end{figure}

From the above, admittedly speculative,  discussion we are lead to the spacetime picture depicted in Figure \ref{figab}.
This picture is reminiscent of the Ashtekar Bojowald paradigm \cite{ab} and of that discussed in \cite{aapersp} wherein gravitational singularities are assumed to be resolved by quantum gravity 
effects, the classical spacetime admits a quantum extension and quantum correlations with earlier thermal Hawking radiation emerge in this quantum extension.
Since, in the spacetime picture of Figure \ref{figab} the quantum extended spacetime arena  is exactly the Minkowskian half plane,  we do expect the quantum state
at its $\I^+$ to be pure. However it is not clear if the state lies in the same Hilbert space ${\cal H}_{Fock}$ as the initial coherent  state on  $\I^-$.

Preliminary calculations suggest that if $\u (x^-)$ is sufficiently smooth and approaches $x^-$ as $x^-\rightarrow \pm\infty$ sufficiently fast
%
the relevant Bogoliuibov transformation between freely falling modes at $\I^-$  and $\I^+$ suffers from no ultraviolet divergences.
There seem, however, to be infra-red divergences. Infrared divergences are typical of massless field theory in 1+1 dimensions and require a more careful treatment 
\cite{glimmjaffe}.  As   indicated in \cite{charlieme}, it is possible that such a treatment may lead to the conclusion that it is only 
ultraviolet divergences which are an obstruction to the unitary implementability of the Bogoliubov transformation. If so, we would expect that under the above 
conditions on $\u$, not only
is the quantum state on $\I^+$ pure, it is also in the same Hilbert space ${\cal H}_{Fock}$ as the initial coherent  state on  $\I^-$.
Note that if $a= 1$ in (\ref{ilong}), then provided $\u (x^-)$ is sufficiently smooth and approaches $x^-$ as $x^-\rightarrow -\infty$ sufficiently fast,
the above discussion applies. For the case $a\neq 1$ we are unable to make any statement and we leave this case (as well as a confirmation of our preliminary calculations 
for the $a=1$ case) for future work. If the state at $\I^+$ is not in ${\cal H}_{Fock}$, we may still interpet it  as an algebraic (and presumably pure) state from the 
perspective of the algebraic approach to quantum field theory \cite{waldbook}.

\section{\label{sec7} Discussion}

In light of the discussion of section \ref{sec6}, we expect that the state at (the quantum extended) $\I^+$ is pure. We expect that at early times, the state at $\I^+$ is
a mixed state of slowly increasing Hawking temperature. Hence it is of interest to understand how this  state is purified to one on extended $\I^+$.
Directly relevant tools to explore this question have been developed in the recent beautiful work of Agullo, Calizaya Cabrera and Elizaga Navascues \cite{ivanetal}.
Envisaged work consists in a putative application of their work to the context of the system studied in this paper.
Another question of physical interest concerns the classically/semiclassically singular region.
While the quantum fluctuations of geometry are expected to be large in this
region, it might still be possible to describe the expectation value geometry through an effective metric. In this regard, the setting for the arguments of \cite{ori}
seem to be satisfied so that it might be possible to argue that the semiclassical singularity is mild in the sense that the conformal factor is
continuous at this singularity. It might then be possible to continue it past the singularity along the lines of \cite{ori} and compare the resulting geometry
to existing proposals in the literature such as \cite{carlowh,aapersp,hay,frol}.

It would also be of interest to understand the semiclassical solution numerically, especially with regard to the behavior of the spacetime geometry and stress energy along the 
last set of rays from the intersection of the marginally trapped tube and the singularity to $\I^+$. In the closely related  semiclassical theory of the 2d CGHS 
model, there is a last ray from exactly such an intersection and extremely interesting universality in quantities such as the back reaction corrected Bondi mass and Bondi flux (scaled down by $N$)
at the last ray \cite{franz}. This universality holds if the black hole formed by collapse is  sufficiently large in a precisely defined sense.
An investigation of physics and possible universality along the last set of rays in the system studied in this work is even more interesting given that, in contrast to
the CGHS case in which the Hawking temperature is independent of mass, the Hawking temperrature here has the standard inverse mass dependence. As far as we can discern,
the black holes studied by Parentani and Piran \cite{pp} are microscopic; it would be exciting if the turn over to less singular behavior in the flux near the last rays
seen by them holds also for initially large black holes.

As emphasized in section \ref{action} a  crucial ingredient in our study of spherical collapse here is the novel `conformal' matter coupling. Besides the 
work by Lowe \cite{lowe} and Parentani and Piran \cite{pp} discussed in section \ref{secpplowe},  such a coupling 
has also been considered by other authors in contexts/approximations different from ours. Specifically, References \cite{pepe-mik,mikvoja,mik,mikvojamaja} consider
a classical action identical to ours but are interested in quantum effects captured by the 1 loop effective action using techniques different from ours.
Reference \cite{pepe} uses a matter coupling identical to ours in a first guess
for subsequent analysis of aspects of the stress energy expectation value in the minimally coupled case;
a key difference stemming from  motivations different from ours, is 
the focus  on staticity and the choice of the Boulware state. Reference \cite{frol} uses the stress energy expectation value for a matter coupling 
identical to ours  as an 
ingredient in its solution for a non-singular, evaporating black hole.

We end on a brief technical note related to the residual choice of conformal coordinates discussed in section \ref{sec2}. Recall that this choice was that of identical
translations in $x^{\pm}$ i.e. $x_{new}^{\pm}\rightarrow x^{\pm} +c$. Clearly, this choice is a special case of Poincare transformations in $x^{\pm}$ and does not affect the identification of 
the vacuum state at $\I^-$. This, in turn, implies that the form of the semiclassical equations remain unchanged. The reader may verify that the asymptotic analysis
of section \ref{sec5} is also unchanged because $\I^+$ is still located at positive infinity of the  new `$+$'
coordinate and the subsequent analysis only depends on derivatives
with respect to the new conformal coordinates, these derivatives being unchanged by virtue of $\frac{dx_{new}^{\pm}}{dx^{\pm}}=1$. Hence as claimed in section \ref{sec2}, the 
physics is independent of this translational ambiguity in the choice of conformal coordinates.

\section*{Acknowledgments} I thank Ingemar Bengtsson for discussions regarding prompt collapse, Kartik Prabhu and Marc Geiller for
discussions regarding asymptotics and  Fernando Barbero for discussions on invertibility of Fourier transforms. I thank Abhay Ashtekar for 
his careful reading of,  and detailed comments on, a draft version of this work.  I thank Fernando Barbero for his kind help with figures.

\section*{Appendices}

\appendix

\section{\label{secaxis}Comments on Axis Boundary Conditions}

Note that the geometry in the vicinity of the axis is, by definition of the axis, non-singular.
As shown in the next section \ref{seca1}
the requirement of non-singularity at the axis is quite powerful and constrains the behavior of $\rho, R$ at the axis as folows:
\be
R=0\;\;\;\;\;\;  \partial_xR =e^{\rho} \;\;\;\;\;\;\;\;\;  \partial_x\rho=0
\label{axiscond}
\ee
\be
\partial_x^2R=0
\label{ignorable}
\ee
In order to obtain these results we assume, in addition to the requirement of non-singular geometry near the axis, that $x^{\pm}$ is a good coordinate system for the 2 geometry defined by $\g2$. Specifically we assume that:\\
(a) The coordinate vector fields $\big(\frac{\partial }{\partial x^{\pm}}\big)^a$ are well behaved 
everywhere and in particular near and at the axis.\\
(b) The conformal factor $e^{2\rho}$ is finite and non-vanishing \\
(c) In the timelike distant past (i.e as $t\rightarrow -\infty$), the metric is flat with $\rho\rightarrow 0$,
$R\rightarrow x$.\\

In section \ref{seca1}  we interpret the requirement of non-singular axial geometry as the finiteness of $\4R,\2R$ and  $G_{ab}v^aw^b$ (for all well behaved vector fields $v,w$).
It is possible that additional conditions are implied by a similar requirement of finiteness of the Weyl tensor. We leave the relevant analysis for future work.
%
%

Due to the timelike nature of the axis, we are not sure if the entire set of conditions (\ref{axiscond})- (\ref{ignorable}) can be consistently imposed.
More in detail, from the point of view of well posed-ness, we have a system of 2nd order differential equations subject to `initial'
conditions at $\I^-$ which is a {\em null} boundary,  as well  boundary conditions at the axis 
at $x=0$,  which is a {\em timelike} boundary.  Issues related to existence and uniqueness
of solutions to such a `mixed' boundary value problem are beyond our expertise and we lack clarity on a number of points.
%
Since the dynamical equations (\ref{gangle})- (\ref{gppmm}) are just the Einstein equations in spherical symmetry,  
the Bianchi identities imply that not all components of these equations are independent. It is not clear to us which of these equations
we should consider as constraints and which as `evolution' equations. It  is also not clear if the conditions (\ref{axiscond}), (\ref{ignorable}) over-constrain the system
and need to be relaxed or if certain of them should be dropped and augmented differently.
Since we are concerned with 4d
spacetime geometry, it is not clear if we should demand axis finiteness of the 2d scalar curvature as above or if this (or different conditions)
would result from
a demand of finiteness of other 4d curvature invariants/components such as those constructed from the Weyl tensor.\\


Instead of explicitly demanding axis finiteness of various physical quantities as in section \ref{seca1} below, 
one may, instead,  adopt a purely differential equation based point of view in which one specifies data $f,\rho,R$ 
which satisfy the `$--$' constraint (\ref{gppmm}) (or (\ref{gppmms})) 
at $\I^-$,  as well as data for $\rho, R$ at the axis $x=0$ such that in the region between the 
axis and $\I^-$ where  the dynamical equations are well defined, a unique solution results.
This is a weaker requirement than the  axial non-singularity as interpreted above. 
A preliminary analysis of the equations suggests that imposition of  the conditions $R=0, \partial_x\rho=0$ at $x=0$ for all $t$ may suffice. We leave a 
detailed analysis and possible confirmation to future work. Note that if indeed these are the correct conditions, the classical and semiclassical solutions
we have constructed in section \ref{sec3} and  proposed in section \ref{sec4} are unique given the initial data $f$ subject to (\ref{aflat}), (\ref{pc}). 



\subsection{\label{seca1}Derivation of (\ref{axiscond}) - (\ref{ignorable}) from Assumptions (a)-(c)}
In what follows we refer to Assumptions (a)-(c) above as A(a)-A(c).
We interpret the requirement that geometry be non-singular at the axis to mean that
the  4d scalar curvature $\4R$, the 2d scalar curvature $\2R$, and  $G_{ab}v^aw^b$ for all well behaved vector fields $v^a, w^b$  are  finite in a small
enough neighborhood of every point on the axis. 
A(a)
  then implies that 
the $\pm$ components of the 4d Einstein tensor $G_{ab}$
at the axis are finite.
In addition, note that the angular killing fields can be rescaled by factors of $R^{-1}$ so as to 
render them of unit norm. These unit norm vector fields, denoted here by $\hat{\Omega}^a$  can be taken to correspond to well defined unit vector fields at  the axis so that $G_{\hat{\Omega}\hat{\Omega}}:= G_{ab} {\hat{\Omega}}^a{\hat{\Omega}}^b$
is also finite (as an example choose $\hat{\Omega}^a = R^{-1}(\frac{\partial}{\partial \theta})^a$ at 
$\theta=0$ with $\phi, \theta$ being the standard polar coordinates on the unit sphere; in cartesian coordinates $(X,Y,Z)$ with $X^2+Y^2+Z^2=R^2$, this corresponds to the unit vector in the `$Z$' direction and clearly admits a well  defined limit at the axis).

To summarise: We have that $R(t,x=0)=0$ so that all derivatives of $R$ with respect to $t$ vanish  at the axis i.e.
\be
(\frac{d}{dt})^m R|_{x=0,t}= 0 \forall m=1,2,3.. \;\;,
\label{dRdt=0}
\ee
and further that  $\4R, \2R, G_{\hat{\Omega}\hat{\Omega}}, G_{\pm \pm}$ and  $G_{+-}$ are finite
at the axis. 

Straightforward computation yields:
\be
\4R= {\frac{1}{R^2}(2+ 8e^{-2\rho} \partial_-R\partial_+R)} - {\frac{1}{R}(16\partial_+\partial_-R)} + {\2R},
\label{4r}
\ee
\be
G_{\hat{\Omega}\hat{\Omega}} = -\frac{\2R}{2} - 4e^{-2\rho}{\frac{1}{R}(\partial_+\partial_-R)}
\label{Gangle}
\ee
\be
G_{\pm \pm}= -\frac{2}{R}( \partial_{\pm}^2R- 2\partial_{\pm}\rho \partial_{\pm}R)
\label{Gppmm}
\ee
Finiteness of $G_{\hat{\Omega}\hat{\Omega}}, \2R $ at the axis together with A(b), equation (\ref{Gangle}) and
(\ref{dRdt=0}) implies that at the axis:
\be
R^{-1}\partial_+\partial_-R=  {\rm finite} \Rightarrow
\partial_+\partial_-R=-(\partial_x)^2R=0
\label{dx2R=0}
\ee
This together with axis finiteness of $\4R, \2R$ implies that 
\be
(2+ 8e^{-2\rho} \partial_-R\partial_+R)=0 \Rightarrow  (\partial_xR)^2= e^{2\rho}
\label{dxR1}
\ee
A(b)  together with (\ref{Gppmm}), the axis finiteness of $G_{\pm\pm}$, (\ref{dRdt=0}), (\ref{dxR1}) imply the finiteness of $\partial_{\pm}^2R$. Equations (\ref{dRdt=0})
and (\ref{dx2R=0}) then imply finiteness of $\partial_t\partial_x R$ at the axis. This implies
that $\partial_xR$ is continuous along the axis so that from (\ref{dxR1}) we have that at the axis:
\be
\partial_xR = e^{\rho}
\label{dxR}
\ee
where we have used assumption A(c) that in the distant past the 4 metric is almost flat so that 
$R \sim x, \rho \sim 0$.  Equation (\ref{dxR}) together with (\ref{Gppmm}), the axis finiteness of $G_{\pm\pm}$, (\ref{dRdt=0})
and (\ref{dx2R=0}) imply that:
\be
\partial_t\partial_x R = 2 \partial_+e^{\rho} = 2\partial_-e^{\rho}
\label{dtdxR2}
\ee
which implies that 
\be
(\partial_+- \partial_-)\rho = 0 
\Rightarrow \partial_x \rho =0
\label{dxrho=0}
\ee

\section{\label{secaft}Coherent states for prompt collapse}

From (\ref{fpm}), (\ref{fp=fm}) and the fact that $f_+$ is of compact support in $x^+$, it follows that at $\I^-$:
\be
f (x^+, x^-=-\infty)= { f}_{({{+}})}(x^+)= \int_{-\infty}^{\infty} dk {\tilde f}_{(+)}(k) \frac{e^{-ikx^+}}{\sqrt{2\pi}}.
\label{ft1}
\ee
Reality of $ f_{  (+) }(x^+)$ implies that ${\tilde f}_{(+)}(k)= {\tilde f}_{(+)}(-k)$. Since ${ f}_{({{+}})}(x^+)$ is continuous and of compact support,
it is absolutely integrable. Hence its Fourier transform ${\tilde f}_{(+)}(k)$ exists and is continuous \cite{tmarsh}.  Let us further restrict attention to ${ f}_{({{+}})}(x^+)$ which is of bounded variation (i.e. 
it is expressible 
as the difference of two bounded, monotonic increasing functions). For such functions the Fourier transform is invertible \cite{tmarsh} and we can reconstruct
${ f}_{({{+}})}(x^+)$ from (\ref{ft1}) with 
\be
{\tilde f}_{(+)}(k) = \int_{-\infty}^{\infty} dk  f_{(+)}(x) \frac{e^{+ikx^+}}{\sqrt{2\pi}}
\label{ft2}
\ee
Defining 
\be
c(k)= \sqrt{2k} {\tilde f}_{(+)}(k) \;\; k\geq 0 ,
\label{ft3}
\ee
we define the coherent state  $|f\ket$ patterned on the function $f$ through:
\be
{\hat a}(k)|f\ket = c(k) |f\ket
\label{ft4}
\ee

We note here that:
\ba 
\lim_{x^+\rightarrow (x^+_i)^+}\frac{m(x^+)}{x^+ -x^+_i} &=& 
\lim_{x^+\rightarrow (x^+_i)^+}\frac{m(x^+)- 0}{x^+-x^+_i} \\
&=& \lim_{x^+\rightarrow (x^+_i)^+} \frac{dm(x^+)}{dx^+}\\
&=&\frac{1}{2}   \lim_{x^+\rightarrow (x^+_i)^+ }   (\partial_+ f(x^+, x^-))^2
\label{ft5}
\ea
where in last line we used (\ref{defm}). Condition (\ref{pc}) together with (\ref{fpm}) then implies that 
\be
\lim_{x^+\rightarrow (x^+_i)^+} \partial_+ f_{(+)}(x^+) > \pm \frac{1}{2\sqrt{2}}
\ee
which indicates a discontinuity in this first derivative at $x^+=x^+_i$ from zero to a non-zero value in accordance with the inequality.
It is evident that there is a rich family of   functions $f_{(+)}$ of this type which are also  continuous functions of compact support and bounded variation.

\section{\label{secaflux}Calculation of Hawking flux for Vaidya spacetime}

The Vaidya line element is given by (\ref{efds2}). As seen in section \ref{sec3}, the coordinate $v$ is identical with the coordinate $x^+$.
However for easy comparision with $(\ref{efds2})$, in this section we will use the notation $v$ instead of $x^+$.

At $\I^+$, $v,R\rightarrow \infty$.  The collapsing matter is compactly supported at $\I^-$. Let its support be  between $v=v_i$ and $v=v_f$.
For $v>v_f$ the spacetime (\ref{efds2}) is Schwarzschild with $v$ being the ingoing Eddington Finkelstein coordinate and 
 $m(v)$ equal to the ADM Mass $M$.
It follows that with 
\be
\u := v-2R^{*},
\label{h1}
\ee
with  $R^*$ the tortoise coordinate:
\be 
R^*:= R+ \frac{R}{2M} \ln (\frac{R}{2M} -1),
\label{h2}
\ee
the line element takes the outgoing Vaidya form (\ref{efout}) with $m_B:=M$. Since we only have infalling classical matter in 
the Vaidya spacetime, from (\ref{balance2}) the stress energy expectation value is given by purely by the quantum `vacuum fluctuation' contribution:
\be
-4\pi R^2 \bra {\hat T}_{\bar u\bar u}\ket 
=  \frac{N\hbar}{48\pi}(\frac{1}{ {\bar u}^{\prime} })^2[ \frac{3}{2}(\frac{{\bar u}^{\prime \prime}}{ {\bar u}^{\prime} })^2 
- \frac{ {\bar u}^{\prime\prime\prime}   }{  {\bar u}^{\prime}  }] \label{h3}
\ee
It remains to compute derivatives of  ${\bar u}$ with respect to $u$. To obtain the Hawking flux, we are interested in computing these derivatives
as $\u \rightarrow \infty$. Since $\u$ is only a function of $u$ and not of $v$, we can compute these
derivatives at any fixed value of $v >v_f$. Let this value be $v=v_0$.  From (\ref{h1}), (\ref{h2}), we have that $\u\rightarrow \infty$ as $R\rightarrow 2M$
i.e. as we approach the horizon along the null line at fixed $v_0$.  Let the value of $u$ at the horizon be $u=u_H$. Since $u$ is a good coordinate,
we have that the conformal factor $e^{2\rho}$ is finite for $u$ near and at $u=u_H$ at fixed $v=v_0$.
Using (\ref{h1}), (\ref{h2}) and (\ref{iv2}), we have that at fixed $v=v_0$:
\be
\frac{R,_u}{R,_{\u}} = \frac{\alpha (u, v_0)}{1-\frac{2M}{R(u,v_0)}} 
\label{h4}
\ee
where we have set $e^{2\rho (u,v_0)}:= \alpha (u, v_0)$. From the fact that $\u (u)$ is independent of $v$, we have that for all $v>v_0$ that 
\be
\u^{\prime} = \frac{\alpha (u, v_0)}{1-\frac{2M}{R(u,v_0)}} .
\label{h5}
\ee
As remarked above, we are interested in late times at $\I^+$ and hence in the behavior of (\ref{h3}) as 
\be 
u\rightarrow u_H \equiv \u\rightarrow \infty \equiv R\rightarrow 2M
\label{h6}
\ee
A long but straightforward calculation shows that in this limit:
\ba
4\pi R^2 \bra {\hat T}_{\bar u\bar u}\ket  &=& -\frac{N\hbar}{48\pi}(\frac{1}{ {\bar u}^{\prime} })^2[ \frac{3}{2}(\frac{{\bar u}^{\prime \prime}}{ {\bar u}^{\prime} })^2 
- \frac{ {\bar u}^{\prime\prime\prime}   }{  {\bar u}^{\prime}  }]  \label{h7}\\
&=&  \frac{N\hbar}{48\pi}\frac{1}{32M^2}
\label{h8}
\ea
We can also compute, in this limit,  a  `corrected' stress energy tensor through the right hand side of (\ref{balance5}):
\be
4\pi R^2 \bra {\hat T}_{\bar u\bar u, corrected}\ket = 
\frac{N\hbar}{96\pi}\frac{({\bar u}^{\prime \prime})^2}{( {\bar u}^{\prime})^4 }.
\label{h9}
\ee
It is straightforward to check that the evaluation of the right hand side of (\ref{h9}) exactly agrees with that of 
(\ref{h8}), that is to say that $ 4\pi R^2\bra {\hat T}_{\bar u\bar u,corrected }\ket = 4\pi R^2 \bra {\hat T}_{\bar u\bar u}\ket$
at late times near $\I^+$. Finally, we may also compute the back reaction corrected Bondi mass at late times on $\I^+$ of the Vaidya spacetime. It  evaluates,
through (\ref{mcorrect}), to
\be
m_{B,initial-corrected}= M + \frac{N\hbar}{48\pi}\frac{1}{4M}
\label{h10}
\ee


\end{document}